\documentstyle[preprint,tighten,prd,aps,epsf,rotate]{revtex}

\begin{document}
\draft

\def\bra#1{{\langle#1\vert}}
\def\ket#1{{\vert#1\rangle}}
\def\coeff#1#2{{\scriptstyle{#1\over #2}}}
\def\undertext#1{{$\underline{\hbox{#1}}$}}
\def\hcal#1{{\hbox{\cal #1}}}
\def\sst#1{{\scriptscriptstyle #1}}
\def\eexp#1{{\hbox{e}^{#1}}}
\def\rbra#1{{\langle #1 \vert\!\vert}}
\def\rket#1{{\vert\!\vert #1\rangle}}
\def\lsim{{ <\atop\sim}}
\def\gsim{{ >\atop\sim}}
\def\nubar{{\bar\nu}}
\def\psibar{{\bar\psi}}
\def\Gmu{{G_\mu}}
\def\alr{{A_\sst{LR}}}
\def\wpv{{W^\sst{PV}}}
\def\evec{{\vec e}}
\def\notq{{\not\! q}}
\def\notk{{\not\! k}}
\def\notp{{\not\! p}}
\def\notpp{{\not\! p'}}
\def\notder{{\not\! \partial}}
\def\notcder{{\not\!\! D}}
\def\notA{{\not\!\! A}}
\def\notv{{\not\!\! v}}
\def\Jem{{J_\mu^{em}}}
\def\Jana{{J_{\mu 5}^{anapole}}}
\def\nue{{\nu_e}}
\def\mn{{m_\sst{N}}}
\def\mns{{m^2_\sst{N}}}
\def\me{{m_e}}
\def\mes{{m^2_e}}
\def\mq{{m_q}}
\def\mqs{{m_q^2}}
\def\mz{{M_\sst{Z}}}
\def\mzs{{M^2_\sst{Z}}}
\def\ubar{{\bar u}}
\def\dbar{{\bar d}}
\def\sbar{{\bar s}}
\def\qbar{{\bar q}}
\def\sstw{{\sin^2\theta_\sst{W}}}
\def\gv{{g_\sst{V}}}
\def\ga{{g_\sst{A}}}
\def\pv{{\vec p}}
\def\pvs{{{\vec p}^{\>2}}}
\def\ppv{{{\vec p}^{\>\prime}}}
\def\ppvs{{{\vec p}^{\>\prime\>2}}}
\def\qv{{\vec q}}
\def\qvs{{{\vec q}^{\>2}}}
\def\xv{{\vec x}}
\def\xpv{{{\vec x}^{\>\prime}}}
\def\yv{{\vec y}}
\def\tauv{{\vec\tau}}
\def\sigv{{\vec\sigma}}
\def\sst#1{{\scriptscriptstyle #1}}
\def\gpnn{{g_{\sst{NN}\pi}}}
\def\grnn{{g_{\sst{NN}\rho}}}
\def\gnnm{{g_\sst{NNM}}}
\def\hnnm{{h_\sst{NNM}}}

\def\xivz{{\xi_\sst{V}^{(0)}}}
\def\xivt{{\xi_\sst{V}^{(3)}}}
\def\xive{{\xi_\sst{V}^{(8)}}}
\def\xiaz{{\xi_\sst{A}^{(0)}}}
\def\xiat{{\xi_\sst{A}^{(3)}}}
\def\xiae{{\xi_\sst{A}^{(8)}}}
\def\xivtez{{\xi_\sst{V}^{T=0}}}
\def\xivteo{{\xi_\sst{V}^{T=1}}}
\def\xiatez{{\xi_\sst{A}^{T=0}}}
\def\xiateo{{\xi_\sst{A}^{T=1}}}
\def\xiva{{\xi_\sst{V,A}}}

\def\rvz{{R_\sst{V}^{(0)}}}
\def\rvt{{R_\sst{V}^{(3)}}}
\def\rve{{R_\sst{V}^{(8)}}}
\def\raz{{R_\sst{A}^{(0)}}}
\def\rat{{R_\sst{A}^{(3)}}}
\def\rae{{R_\sst{A}^{(8)}}}
\def\rvtez{{R_\sst{V}^{T=0}}}
\def\rvteo{{R_\sst{V}^{T=1}}}
\def\ratez{{R_\sst{A}^{T=0}}}
\def\rateo{{R_\sst{A}^{T=1}}}

\def\mro{{m_\rho}}
\def\mks{{m_\sst{K}^2}}
\def\mpi{{m_\pi}}
\def\mpis{{m_\pi^2}}
\def\mom{{m_\omega}}
\def\mphi{{m_\phi}}
\def\Qhat{{\hat Q}}

\def\FOS{{F_1^{(s)}}}
\def\FTS{{F_2^{(s)}}}
\def\GAS{{G_\sst{A}^{(s)}}}
\def\GES{{G_\sst{E}^{(s)}}}
\def\GMS{{G_\sst{M}^{(s)}}}
\def\GATEZ{{G_\sst{A}^{\sst{T}=0}}}
\def\GATEO{{G_\sst{A}^{\sst{T}=1}}}
\def\mdax{{M_\sst{A}}}
\def\mustr{{\mu_s}}
\def\rsstr{{r^2_s}}
\def\rhostr{{\rho_s}}
\def\GEG{{G_\sst{E}^\gamma}}
\def\GEZ{{G_\sst{E}^\sst{Z}}}
\def\GMG{{G_\sst{M}^\gamma}}
\def\GMZ{{G_\sst{M}^\sst{Z}}}
\def\GEn{{G_\sst{E}^n}}
\def\GEp{{G_\sst{E}^p}}
\def\GMn{{G_\sst{M}^n}}
\def\GMp{{G_\sst{M}^p}}
\def\GAp{{G_\sst{A}^p}}
\def\GAn{{G_\sst{A}^n}}
\def\GA{{G_\sst{A}}}
\def\GETEZ{{G_\sst{E}^{\sst{T}=0}}}
\def\GETEO{{G_\sst{E}^{\sst{T}=1}}}
\def\GMTEZ{{G_\sst{M}^{\sst{T}=0}}}
\def\GMTEO{{G_\sst{M}^{\sst{T}=1}}}
\def\lamd{{\lambda_\sst{D}^\sst{V}}}
\def\lamn{{\lambda_n}}
\def\lams{{\lambda_\sst{E}^{(s)}}}
\def\bvz{{\beta_\sst{V}^0}}
\def\bvo{{\beta_\sst{V}^1}}
\def\Gdip{{G_\sst{D}^\sst{V}}}
\def\GdipA{{G_\sst{D}^\sst{A}}}
\def\fks{{F_\sst{K}^{(s)}}}
\def\FIS{{F_i^{(s)}}}
\def\fpi{{F_\pi}}
\def\fk{{F_\sst{K}}}

\def\RAp{{R_\sst{A}^p}}
\def\RAn{{R_\sst{A}^n}}
\def\RVp{{R_\sst{V}^p}}
\def\RVn{{R_\sst{V}^n}}
\def\rva{{R_\sst{V,A}}}
\def\xbb{{x_B}}

\def\PR#1{{{\em   Phys. Rev.} {\bf #1} }}
\def\PRC#1{{{\em   Phys. Rev.} {\bf C#1} }}
\def\PRD#1{{{\em   Phys. Rev.} {\bf D#1} }}
\def\PRL#1{{{\em   Phys. Rev. Lett.} {\bf #1} }}
\def\NPA#1{{{\em   Nucl. Phys.} {\bf A#1} }}
\def\NPB#1{{{\em   Nucl. Phys.} {\bf B#1} }}
\def\AoP#1{{{\em   Ann. of Phys.} {\bf #1} }}
\def\PRp#1{{{\em   Phys. Reports} {\bf #1} }}
\def\PLB#1{{{\em   Phys. Lett.} {\bf B#1} }}
\def\ZPA#1{{{\em   Z. f\"ur Phys.} {\bf A#1} }}
\def\ZPC#1{{{\em   Z. f\"ur Phys.} {\bf C#1} }}
\def\etal{{{\em   et al.}}}

\def\delalr{{{delta\alr\over\alr}}}
\def\pbar{{\bar{p}}}
\def\lamchi{{\Lambda_\chi}}

\title{
\hfill{\normalsize INT \#DOE/ER/40561-291-INT96-00-149}\\
\hfill{\normalsize MKPH-T-96-25}\\[2.5cm]
Nucleon Strangeness and Unitarity}

\author{M.J. Musolf$^{a,}$\thanks{National Science Foundation
Young Investigator}, H.-W. Hammer$^{a,b}$, and 
D. Drechsel$^{b}$\\[0.3cm]
}
\address{
$^a$ Institute for Nuclear Theory, 
University of Washington, Seattle, WA 98195 USA\\
$^b$ Universit{\"a}t Mainz, Institut f{\"u}r Kernphysik,
J.-J.-Becher Weg 45, D-55099 Mainz, Germany
}


\maketitle

\begin{abstract}
The strange-quark vector current form factors of the nucleon are analyzed
within the framework of dispersion relations. Particular attention is paid to
contributions made by $K\bar{K}$ intermediate states to the form factor
spectral functions. It is shown that, when the $K\bar{K}\to N\bar{N}$
amplitude is evaluated in the Born approximation, the 
$K\bar{K}$ contributions
are identical to those arising from a one-loop calculation and entail
a serious violation of unitarity. The mean square strangeness radius and
magnetic moment are evaluated by imposing unitarity bounds on 
the kaon-nucleon
partial wave amplitudes. The impact of including the kaon's 
strangeness vector current
form factor in the dispersion integrals is also evaluated. 
\end{abstract}


\narrowtext

\section{Introduction}
\label{sec:intro}
The low-energy structure of the nucleon's $s\bar{s}$ sea has become a topic
of serious study in the hadron structure community \cite{Mus94a}. 
While deep inelastic
scattering (DIS) has provided information about the light-cone
momentum distribution of the strange sea \cite{CCF93}, little is known about
the corresponding spatial and spin distributions or about the role played
by the sea in the nucleon's response to a low-energy probe. In an effort
to study some of these low-energy characteristics of the sea, several
semi-leptonic scattering experiments are underway and/or planned at
MIT-Bates, TJNAF (formerly CEBAF), MAMI, and LANL. Parity-violating 
experiments
using polarized electrons \cite{Mck89,Pit94,Bec91,Bei91,Fin91,Har93}
are aimed primarily at probing nucleon
matrix elements of the strange-quark vector current, which is parametrized
by the strangeness electric and magnetic form factors, $\GES$ and $\GMS$,
respectively. Additionally, one expects the neutrino scattering data 
from LANL \cite{Lou89} to yield
new limits on the strange-quark axial vector matrix element, characterized
by the axial form factor $\GAS$.

The corresponding problem for hadron structure theory is to
compute these form factors and their leading moments, which depend crucially
on non-perturbative aspects of QCD, in a credible manner.  To this end,
one may choose from a number of different strategies, each with its 
particular merits and limitations:
\begin{enumerate}
\item[(a)] Lattice QCD. To date, lattice calculations of the 
strangeness axial charge $\Delta s = \GAS(0)$ \cite{Liu95}
and strangeness magnetic moment $\mu^s=\GMS(0)$ \cite{Lei96}
have been carried out in the quenched approximation. The results for
$\Delta s$ are essentially consistent with the experimental value extracted
from polarized DIS measurements \cite{Ash89}. The first lattice results for 
$\mu_s$, however, differ in sign from the preliminary experimental
value obtained by the SAMPLE collaboration \cite{Bei96}.  
With the future development
of more sophisticated lattice methods, one would anticipate better agreement 
between calculated and experimental values for these strangeness moments. 
The primary attraction of lattice calculations is that they provide the
most direct, first principles, non-perturbative computations using QCD. 
By themselves, however, they may not provide as much insight as one would 
like into the mechanisms which govern the sign and scale of the strangeness 
form factors. Moreover, obtaining results for the non-leading 
$Q^2$-dependence
of the form factors may prove to be a formidable task.
\item[(b)] Effective Theory. A complementary approach is to work with
effective hadronic degrees of freedom rather than the quark and gluons of
QCD, incorporating the underlying symmetries of the QCD Lagrangian into the
effective hadronic Lagrangian. This approach, in the guise of chiral
perturbation theory (CHPT), has seen considerable success in a variety of
contexts \cite{Don92}. 
A particular advantage of CHPT is its reliance on chiral symmetry
and existing data, rather than on microscopic calculations, to determine
quantities (chiral counterterms) whose values reflect the impact of 
short-distance hadronic interactions. Moreover, CHPT provides one with a 
useful language  in which to describe the strong interaction dynamics 
responsible for magnitude and sign of a particular quantity.
In the case of the strangeness vector current form factors, however,
CHPT cannot be used to make a model-independent prediction, as discussed  
in detail in Ref. \cite{Mus96}.
\item[(c)] Hadronic Models. A variety of model calculations for 
the strangeness form factors have been carried out 
\cite{Jaf89,Ham96,Mu94b,Koe92,For94,Coh93,Par91,Hon93,Pha94,Ito95,Gei96}, 
among which there
appears little consensus as to the magnitude or sign of the different
strangeness moments. Some models start from the effective theory framework
and invoke additional, model assumptions in order to arrive at predictions.
Others, such as the cloudy bag model or non-relativistic quark model, 
attempt to provide a more microscopic description of the form factors. 
The appeal of models is that they attempt to incorporate one's 
intuition about the physics which drives a particular aspect of 
hadron structure. Nevertheless,
the correspondence between any model and the dynamics of QCD is open to 
debate. In the case of nucleon strangeness, this situation is reflected in
the wide range of model predictions for strangeness form factors. If one
wishes to understand the spin and spatial distribution of the $s\bar{s}$ sea
in terms of QCD, then models would appear to have a limited usefulness.
\item[(d)] Dispersion Relations. In the present paper, we turn to this
approach to try and derive insight into the strangeness form factors.
The use of dispersion relations (DR) has several merits, some
of which are similar to those of effective theory. Like
CHPT, DR employ effective, hadronic degrees of freedom rather
than the quarks and gluons of QCD. Similarly, DR offer a 
rigorous and, in principle, model-independent framework 
in which to understand 
the hadronic mechanisms which govern form factors. Both approaches attempt
to relate experimental hadronic amplitudes to the form factors of interest,
relying in the one case on chiral symmetry (CHPT) and in the other on
analyticity and causality (DR). Although DR and CHPT are not QCD in
a microscopic sense, they nevertheless embody QCD insofar as it is
responsible for the experimental strong interaction observables used as 
input for a calculation. 
\item[]For the present purposes, DR offer additional 
advantages not afforded by CHPT. First, ultra-violet divergences can be 
eliminated using unitarity
bounds rather than subtraction constants. In the case of the strangeness
form factors, it is one's inability to determine the finite part of these
counterterms which renders CHPT un-predictive \cite{Mus96}. 
Second, DR can be used to
convert a given body of experimental data into predictions for the behavior
of form factors over a range of momentum transfer. This situation contrasts
with that of CHPT, which involves an expansion in powers of external 
momentum and requires the determination of additional 
counterterms at each order in the expansion. 
The limitations of DR, as an effective hadronic framework,
are essentially set the by the availability of sufficient data on strong
and electroweak amplitudes. In the absence of such available data, one is
forced, within this framework, to resort to ancillary approximations.
\end{enumerate}

The application of DR to the study of nucleon form
factors is not new. Well before the discovery of QCD, DR
were used to analyze the nucleon's electromagnetic (EM) form 
factors \cite{Fed58,Che58,Fra60}.
In addition to shedding light on the nucleon's EM structure, dispersion
relation analyses have allowed one to extract the couplings of various
mesons to the nucleon \cite{Hoe76,Mer96}. More recently, DR have been
employed to make predictions for the nucleon's strange-quark vector
current form factors \cite{Jaf89,Ham96,Mus96}. 
These predictions have generally invoked the 
assumption of vector meson dominance, which, based on experience with
the nucleon's isovector EM form factor as well as on general grounds, is
debatable. In principle, any nucleon form factor receives both resonant
and non-resonant (continuum) contributions. In the case of the nucleon's
isovector EM charge radius, for example, the continuum contribution is
non-negligible. While one can make a case for resonance dominance in the
case of the nucleon's mean square strangeness radius based on
a model-dependent extension of the effective theory approach 
\cite{Mus96}, the logic rests on untested assumptions about the continuum 
contributions. Indeed, arriving at a rigorous, consistent, and
model-independent analysis which incorporates both continuum and resonance
contributions to the strangeness form factors remains an open problem for
effective hadronic approaches.

With this problem in mind, we focus on the behavior of the multi-meson
continuum, emphasizing in particular the two-kaon contribution. The
continuum contribution has been studied previously, with both CHPT and
models, using one-loop kaon-strange baryon ($B$) calculations 
\cite{Mus96,Mu94b,Koe92,For94,Coh93,Ito95}. In the 
$t$-channel, such loops represent approximations to the $K\bar{K}$ and 
$B\bar{B}$ intermediate state contributions. 
Although the lightest intermediate
state which can contribute to the form factors contains three pions, the
$KB$ loop calculations have been justified under the {\em ansatz} that
hadronic states having valence $s$ and $\bar{s}$ quarks -- the so-called
\lq\lq kaon cloud" -- should give the
dominant contribution. Using the $K\bar{K}$ intermediate state as an
illustrative example, we show
how one-loop estimates of the continuum contribution can entail a serious
violation of unitarity, and evaluate the bounds on the 
continuum contribution which result from the imposition of unitarity. 
Our results indicate that
effects which go beyond one-loop order -- in effect, kaon rescattering
corrections -- cannot be neglected. We also analyze the impact 
on predictions for the nucleon's strangeness form factors made 
by one's choice for the kaon strangeness form factor, $\fks$.  
We find that this impact is non-trivial. Consequently, 
since $\fks$ has not been measured, one's choice for its form
necessarily introduces a certain degree of model-dependence into the 
dispersion relation analysis. Finally, we note that the conclusions of
the present study are provisional. We are unable to make any rigorous
statements about contributions to the dispersion integrals in the kinematic
regime where unitarity does not apply. In a subsequent paper we will report
on our attempt to estimate these contributions by drawing upon existing
kaon-nucleon scattering data. Similarly, we postpone to a future discussion
any treatment of other multi-meson continuum and baryon intermediate state
contributions. In essence, our study follows the spirit of the analysis
of Ref. \cite{Fed58}. In that work, the impact of unitarity constraints
and inclusion of a pseudoscalar electromagnetic (EM) form factor 
were treated for the $\pi\pi$ contribution to the nucleon's 
isovector EM form factors.

Our discussion of these points is organized as follows. In section II, we
review the dispersion relation formalism as it applies to nucleon form
factors. We also specify this formalism to the two-kaon continuum case,
introducing our own version of the $K\bar{K}$ partial waves to make unitarity
constraints transparent. In section III, we compare the two-kaon 
contribution
in the Born approximation, which is equivalent to a one-loop calculation,
with a calculation which incorporates the unitarity bounds and $\fks$.
In section IV we discuss our results for the mean-square strangeness
radius and magnetic moment. Section V
summarizes our conclusions and is followed by an Appendix.

\section{Formalism}
\label{sec:formal}

\def\bpp{{b_1^{1/2,\,1/2}}}
\def\bpm{{b_1^{1/2,\,-1/2}}}
\def\bll{{b_1^{\lambda_1, \lambda_2}}}

In writing down dispersion relations for the nucleon's strangeness form
factors, we find it useful to follow the treatments of Drell and
Zachariasen \cite{DrZ60} and Federbush, Goldberger, and Treiman \cite{Fed58}. 
We also choose to work with the standard Dirac and Pauli form factors, 
$\FOS$ and $\FTS$, respectively, defined as
\begin{equation}
\label{def_ff}
\bra{N(p')}\bar{s}\gamma_\mu s\ket{N(p)}=\bar{U}(p')\left[\FOS(t)
\gamma_\mu+{i\FTS(t)\over 2\mn}\sigma_{\mu\nu}Q^\nu\right]U(p)\ \ \ ,
\end{equation}
where $U(p)$ is a spinor associated with the nucleon state $\ket{N(p)}$.
Since the nucleon has no net strangeness, one has $\FOS(0)=0$.
The form factors $\FIS$ ($i=1,2$) are related to the Sachs electric
and magnetic form factors \cite{Sac62} via 
\begin{eqnarray}
\label{def_sac}
\GES&=&\FOS-\tau\FTS\ \ ,\\
\GMS&=&\FOS+\FTS\ \ \ ,\nonumber
\end{eqnarray}
where $t=Q^2=(p'-p)^2$, $\tau=-t/4\mns$, and $p$ ($p'$) is the initial
(final) nucleon four momentum. We are
particularly interested in the leading moments associated with the
$\FIS$: the mean square strangeness radius and magnetic moment,
defined as
\begin{eqnarray}
\label{def_rho}
\rho^s_D&=&{d\FOS\over d\tau}\bigg\vert_{\tau=0}\, ,\\
\label{def_mu}
\mu^s&=&\FTS(0) \, .
\end{eqnarray}
We have chosen a dimensionless version of the mean square 
radius, which is related to the corresponding dimensionfull quantity as
\begin{eqnarray}
\langle r_s^2\rangle = 6{d\FOS\over d Q^2}\bigg\vert_{Q^2=0} = -{3\over 2}
\mns\rho^s_D \ \ \ .
\end{eqnarray}                                               

In order to obtain a dispersion relation
for one of the $\FIS(t)$ ($i=1,2$), where $t$ is real, 
one must assume that there exists an analytic continuation $\FIS(z)$
which approaches $\FIS(t)$ as $z\to t+i\epsilon$, which is analytic in
the upper half plane, and which has a branch cut on the real axis for
$t$ greater than some threshold, $t_0$. In addition, one must assume
that 
\begin{equation}
\label{high_t}
{\FIS(z)\over z^n}\rightarrow 0
\end{equation}
as $z\to\infty$ anywhere in the upper half plane for some non-negative
integer $n$. In this case, a straightforward application of Cauchy's
Theorem (using a circular contour excluding the branch cut)
leads to the relations
\begin{equation}
\label{unsub}
\FIS(t)={1\over\pi}\int_{t_0}^\infty\ {\hbox{Im}\  \FIS(t')\over t'-t-
i\epsilon}dt'
\end{equation}
in the case of $n=0$,
\begin{equation}
\label{sub}
\FIS(t)-\FIS(0)={t\over\pi}\int_{t_o}^\infty\ {\hbox{Im}\  \FIS(t')\over
t'(t'-t-i\epsilon)}
\end{equation} 
in the case of $n=1$, and so forth. 

Employing as large a value of $n$
as possible is desirable in order to improve the convergence of the
function $\FIS(z)/z^n$ on the circular part of the contour at infinity.
One has no way of knowing, {\em a priori}, which is the minimum value of
$n$ needed to guarantee that this contribution to the contour integral
vanishes. The appropriate choice therefore remains one of the inherent
uncertainties in the dispersion relation approach. It is conventional to
use a subtracted dispersion relation (Eq. (\ref{sub})) for the 
Dirac form factor ($i=1$), since one knows on 
general grounds that the value of the form
factor at $t=0$ is just the charge associated with the corresponding
current. In the case of $\bra{N(p')}\bar{s}\gamma_\mu s\ket{N(p)}$, one has
$\FOS(0)=0$ since the nucleon carries no net strangeness. In the case of
the magnetic form factor, one would like to predict its value at $t=0$
rather than using it as a subtraction constant. Hence, we use the 
un-subtracted dispersion relation (Eq. (\ref{unsub})) for $\FTS(t)$. 

The essential physics content entering the DR enters
through the spectral functions, $\hbox{Im}\  \FIS(t)$. To analyze these
spectral functions, we follow Refs. \cite{Fed58,DrZ60} and work in the 
$N\bar{N}$ production channel, where the corresponding current matrix 
element is
\begin{equation}
\label{def_ff_t}
\bra{N(p);\bar{N}(\pbar)}\bar{s}\gamma_\mu s\ket{0}=\bar{U}(p)\left[
\FOS{(t)}\gamma_\mu+{i\FTS{(t)}\over 2\mn}\sigma_{\mu\nu}P^\nu
\right]V(\pbar)
\end{equation}
with $P^\mu=(\pbar+p)^\mu$, $t=P^2$, and $V(\pbar)$ being an anti-nucleon 
spinor. In order to obtain the imaginary parts of the $\FIS$, 
we reduce the anti-nucleon
using the LSZ formalism and take the absorptive part. As in Ref.
\cite{Fed58,DrZ60} 
the resulting contribution to the spectral functions arises from 
\begin{eqnarray}
\label{spec_t}
\hbox{Im}\ \bra{N(p);\bar{N}(\pbar)}\bar{s}\gamma_\mu s\ket{0}
&\rightarrow& \\
{\pi\over\sqrt{Z}}(2\pi)^{3/2}&{\cal N}\sum_{n}& 
\bra{N(p)}\bar{J_N}(0)\ket{n}\bra{n}\bar{s}\gamma_\mu s\ket{0} V(\pbar)
\delta^4(p+\pbar-p_n)\ \ \ ,
\nonumber
\end{eqnarray}
where ${\cal N}$ is a nucleon spinor normalization factor, $Z$ is the
nucleon's wavefunction renormalization constant, and $\bar{J_N}(x)=
J_N^{\dag}(x)\gamma_0$ with $J_N(x)$ being a nucleon source satisfying
\begin{equation}
\label{n_sour}
(i\notder - \mn){\hat\psi}_N(x)=J_N(x)
\end{equation}
and with $\hat\psi_N$ being the nucleon field. The content of the spectral
function, as expressed in Eq.~(\ref{spec_t}), has a useful diagrammatic
representation as shown in Fig. 1.

The states $\ket{n}$ of momentum $p_n$ appearing in the sum are stable 
(with respect to the strong interaction). Consequently, no resonances 
appear in the sum -- only asymptotic final states. In addition, the states
$\ket{n}$ must carry the same quantum numbers as the current $\bar{s}
\gamma_\mu s$: $ I^G (J^{PC}) = 0^- (1^{--}) $. Moreover,
owing to the presence of the source $J_N(0)$, they can have
no net baryon number. In the purely mesonic sector, the lightest such
states are $3\pi$, $5\pi$, $7\pi$, $2K$, $9\pi$, $KK\pi$, $\ldots$. In
the case of the baryons, one has $N\bar{N}$, $\Lambda\bar{\Lambda}$,
$\ldots$. One may also consider states containing both mesons and baryons,
such as $N\bar{N}\pi\pi$. From this enumeration of states, and the delta
function appearing in Eq. (\ref{spec_t}), one sees that the first cut 
in the dispersion
integral appears at the three $\pi$ production threshold, $t_0=9m_\pi^2$.
Higher-mass intermediate states generate additional cuts in
the complex plane. 

Many of the predictions for the $\FIS$ reported in the literature are
based on approximations to the spectral functions appearing in Eqs. 
(\ref{unsub}) and (\ref{sub}). In the work of Ref. \cite{Jaf89}, 
and up-dated in Ref. \cite{Ham96}, a vector meson dominance
approximation was employed, which amounts to assuming that one may write
the spectral function as
\begin{equation}
\hbox{Im}\  \FIS(t) = \pi\sum_j a_j\delta(t-m_j^2)
\end{equation}
where \lq\lq $j$" denotes a particular vector meson resonance 
({\em e.g.}, $\omega$,
$\phi$) and where the sum runs over a finite number of resonances. In
terms of the formalism of section II, this approximation omits any explicit
mention of multi-meson intermediate states $\ket{n}$ and assumes that
collectively the products $\bra{N(p)}\bar{J_N}(0)\ket{n}\bra{n}
\bar{s}\gamma_\mu s\ket{0}V(\pbar)$
are strongly peaked in the regions near one or more vector
meson masses. 

In contrast,
a variety of hadronic effective theory and model calculations have focused
on contributions from the two kaon intermediate state 
\cite{Mus96,Mu94b,Koe92,For94,Coh93,Ito95} even though it is
not the lightest state appearing in the sum. The reason is based primarily
on an intuition that kaons, which contain valence $s$ or $\bar{s}$ quarks,
ought to give larger contributions to the matrix element $\bra{n}\bar{s}
\gamma_\mu s\ket{0}$ than a purely pionic state in which there are no
valence $s$ or $\bar{s}$ quarks. The validity of this {\em ansatz} is
open to question for at least two reasons. First,
the three $\pi$ threshold is significantly below the $K\bar{K}$ threshold. 
Consequently, the $3\pi$ contribution will be weighted more strongly
in the dispersion integral than the $K\bar{K}$ contribution
(owing to the denominators in Eqs. (\ref{unsub},\ref{sub})). Second, 
three pions can resonate into a state having the same
quantum numbers as the $\phi$ (nearly pure $s\bar{s}$), and thereby 
generate a non-trival contribution to the current matrix element 
\cite{rljdi}. 
Indeed, the $\phi$ has roughly a 15\% branch to multi-pion 
final states (largely via a $\rho\pi$ resonance). Although no resonances
appear explicitly in the sum over states in Eq. (\ref{spec_t}), 
the impact of resonances
nevertheless enters via the current matrix element $\bra{n}\bar{s}\gamma_\mu
s\ket{0}$ and $N\bar{N}$ production amplitude 
$\bra{N(p)}\bar{J_N}(0)\ket{n}V(\pbar)$. It is noteworthy that the
kaon-cloud predictions for $\rho^s_D$ are typically smaller in magnitude
than the vector meson dominance predictions and have the opposite sign. 

We leave the relative size of the multi-pion and two kaon contributions to
a future study, and focus in the present paper on the two kaon state. In
doing so, our goal is to indicate how one-loop
effective theory and model calculations
which assume two kaon dominance violate unitarity. In addition, we seek to
illustrate the impacts on $\FIS$ predictions made by (a) 
the imposition of unitarity and (b) the inclusion of a form factor in 
the matrix element $\bra{n}\bar{s}\gamma_\mu s\ket{0}$. To that end, we
first decompose the $K\bar{K}\to N\bar{N}$ amplitude into partial waves and
relate them to the form factor spectral functions. We subsequently
discuss possible parametrizations of the kaon strangeness form factor.

\subsection{Spectral Functions, Partial Waves, and Unitarity}

By expanding the $K\bar{K}\to N\bar{N}$ amplitude 
in partial waves, we are able to identify the pieces which 
contribute to the absorptive part of the nucleon current matrix element
(Eq. (\ref{spec_t})) and impose the constraints 
of unitarity in a straightforward
manner. In doing so, it is convenient to follow the helicity
amplitude formalism of Jacob and Wick \cite{JaW59}. 
We correspondingly assign
the nucleon and anti-nucleon helicities $\lambda_1$ and $\lambda_2$,
respectively, and write the corresponding $S$-matrix element as
\begin{eqnarray}
\label{s_norm}
& &\bra{N(p,\lambda_1)\bar{N}(\pbar,\lambda_2)}{\hat S}\ket{K(k_1)K(k_2)}
= \\ & &\hphantom{N(p,\lambda_1)}
 (2\pi)^4\delta^4(p+\pbar-k_1-k_2)(2\pi)^2 \left[{64 t\over t-4 m_K^2}
\right]^{1/2} \bra{\theta, \phi, \lambda_1, \lambda_2}{\hat S}(P)\ket{00}
\nonumber
\end{eqnarray}
where $P=p+\pbar=k_1+k_2$, $t=P^2$, and $m_K$ is the kaon mass. Defining,
\begin{eqnarray}
\label{cmdef}
q_1^\mu&=&\frac{1}{2}(k_1-k_2)^\mu \\
q_2^\mu&=&\frac{1}{2}(\pbar-p)^\mu \nonumber
\end{eqnarray}
we have $(\theta, \phi)$ as the polar and azimuthal angles made by
${\vec q_2}$ with respect to ${\vec q_1}$ (the \lq\lq $\ket{00}$''
indicates that the incoming mesons have no helicities and that 
we have chosen the $z$-axis to be along ${\vec q_1}$). 

Following Ref. \cite{JaW59}, we expand the matrix element 
$\bra{\theta, \phi, 
\lambda_1, \lambda_2}{\hat S}(P)\ket{00}$ in partial waves as
\begin{equation}
\label{par_d}
S_{\lambda_1, \lambda_2} \equiv
\bra{\theta, \phi, \lambda_1, \lambda_2}{\hat S}(P)\ket{00}=\sum_J
\left({2J+1\over 4\pi}\right) b_J^{\lambda_1\, \lambda_2} \;
{\cal D}_{0\mu}^J(\phi, \theta, -\phi)^{\ast}
\end{equation}
where ${\cal D}_{\nu\, \nu'}^J(\alpha, \beta, \gamma)$ is the standard
Wigner rotation matrix, where $\mu=\lambda_1-\lambda_2$ in 
Eq. (\ref{par_d},)
and where the $b_J^{\lambda_1\, \lambda_2}$ define the partial waves of
angular momentum $J$.

Using the above definitions and imposing the requirement of unitarity
on the $S$-matrix,
\begin{equation}
S^{\dag} S = 1
\end{equation}
one has that
\begin{equation}
\label{ubs}
|b_J^{\lambda_1, \lambda_2}|\leq 1
\end{equation}
for $t\geq 4\mns$. 

In the expression for the spectral function appearing in Eq. (\ref{s_norm}), 
only the $J=1$ partial waves appear since the states $\ket{n}$ must carry
the same quantum numbers as the current $\bar{s}\gamma_\mu s$. Moreover,
it is well known that one has only two independent amplitudes for the
scattering reaction $KN\to KN$ and its crossed channel version $K\bar{K}\to
N\bar{N}$. These amplitudes are commonly chosen to be the $A$ and $B$
amplitudes defined by the $T$-matrix element
\begin{equation}
\label{def_t}
T(-\pbar\,\lambda_1, k_1; p\,\lambda_2, -k_2)=\bar{U}(p\lambda_2)\left[
A+\frac{1}{2} B(\notk_1-\notk_2)\right]V(\pbar\lambda_1)
\end{equation}
where we have employed crossing symmetry to obtain the $t$-channel version
of the $KN\to KN$ scattering amplitude. It is a straightforward exercise
to relate the $A$ and $B$ amplitudes to the $b_1^{\lambda_1\lambda_2}$
\cite{Ste96}.
We choose the two independent partial waves to correspond to $(\lambda_1,
\lambda_2)=(\frac{1}{2}, \frac{1}{2})$ and $(\frac{1}{2}, -\frac{1}{2})$.
We obtain
\begin{eqnarray}
\label{par_p}
\bpp&=&-\left({1\over 2\pi}\right)\left({t-4 m_K^2 \over 64 t}
\right)^{1/2}\biggl\{{p\over\mn}\int_{-1}^1 dx\ x A \\
&&-{k\over 2}\int_{-1}^1 dx\ (3x^2-1) B +{k\over 2}\int_{-1}^1 dx\ 
(x^2-1) B\biggr\}\nonumber \\
&&\nonumber \\
\label{par_m}
\bpm&=&\left({1\over 2\pi\sqrt{2}}\right)\left(
{t-4 m_K^2 \over 64 t}
\right)^{1/2}\int_{-1}^1 dx\ \left({Ek\over \mn}\right)(1-x^2)B
\end{eqnarray}
where $x=\cos\theta$, $k=|{\vec k_1}|=|{\vec k_2}|$, and
$p=|{\vec p}|=|{\vec \pbar}|$ in the $N\bar{N}$ cm frame, and $E=
\sqrt{p^2+\mns}$. 

For future reference, we also note the following $N\bar{N}$ production
threshold relation between the partial waves:
\begin{equation}
\label{th_rel}
\bpm=\sqrt{2}\,\bpp
\end{equation}
as $t\to 4\mns$ (or $P\to 0$). The origin of this relation is easy to
understand. Since the $N$ and $\bar{N}$ have opposite intrinsic parities
while the intrinsic parities of the $K$ and $\bar{K}$ are the same, the
spin $\times$ spatial part of the $K\bar{K}\to N\bar{N}$ amplitude must
transform as a pseudoscalar. In the $K\bar{K}$ c.m. frame, one may therefore
write the two independent amplitudes as
\begin{equation}             
\label{sll}
S_{\lambda_1, \lambda_2} = \chi_{\lambda_1}^{\dag}\left[ f_1 
{\vec\sigma}\cdot {\vec k} + f_2 {\vec\sigma}\cdot {\vec p} 
\right]\chi_{\lambda_2}^{\vphantom{\dag}}\ \ \ ,
\end{equation}
where ${\vec k}\equiv{\vec k_1}$ and ${\vec p}\equiv{\vec p_1}$ and
where the functions $f_{i}$ may depend on $k^2$, ${\vec k}\cdot 
{\vec p}$, {\em etc}.
At threshold, one has ${\vec p}=0$, so that only the amplitude proportional
to $f_1$ survives. From Eq. (\ref{sll}) we obtain
\begin{eqnarray}
\label{spp}
S_{1/2, 1/2} &\rightarrow& f_1 k_z =  f_1 k \sqrt{4\pi\over 3}
Y_{10}^{\ast}(\theta, \phi) \\
\label{spm}
S_{1/2, -1/2} &\rightarrow & f_1 (k_x-ik_y)  =  -f_1 k \sqrt{8\pi\over 3}
	Y_{11}^{\ast}(\theta,\phi) \ \ \ 
\end{eqnarray}
at threshold ($f_1$ may now only depend on $k$). 
The partial waves are obtained by inverting Eq. (\ref{par_d}),
yielding
\begin{eqnarray}
\label{pwpp}
\bpp&=&4\pi\sqrt{\pi\over 3}\int_{-1}^1\, d\cos\theta\; 
	Y_{10}(\theta, \phi)S_{1/2, 1/2} \\
\label{pwpm}
\bpm&=&-4\pi\sqrt{\pi\over 3}\int_{-1}^1\, d\cos\theta\; 
  	Y_{11}(\theta, \phi)S_{1/2, -1/2}\ \ \ .
\end{eqnarray}
The foregoing expressions imply that the
$\bll$ are now independent of the angle $\phi$.
Using the orthonormality of the spherical harmonics, one
sees immediately from Eqs. (\ref{spp}-\ref{pwpm}) that the two partial
waves are related at threshold as indicated in Eq. (\ref{th_rel}).

We may now write the $\hbox{Im}\ \FIS(t)$ in terms of the two independent
$\bll$. Starting from the general expression in Eq. (\ref{spec_t}), 
specifying the states
$\ket{n}$ to contain two kaons only, and replacing the sum $\sum_n$ by
appropriate integrals over two kaon phase space, we obtain expressions for
the spectral functions:
\begin{eqnarray}
\label{imf1}
\hbox{Im}\  \FOS(t)&=&\hbox{Re}\left\{\left({\mn Q\over 4 P^2}\right)\left[
{E\over\sqrt{2}\mn}\bpm-\bpp\right]\fks(t)^{\ast}\right\}\\
&& \nonumber \\
\label{imf2}
\hbox{Im}\  \FTS(t)&=&\hbox{Re}\left\{\left({\mn Q\over 4 P^2}\right)\left[
\bpp-{\mn\over\sqrt{2}E}\bpm\right]\fks(t)^{\ast}\right\}
\end{eqnarray}
where
\begin{eqnarray}
P&=&\sqrt{t/4-\mns}\\
Q&=&\sqrt{t/4-m_K^2}\ \ \ .
\end{eqnarray}
The kaon's strangeness form factor, $\fks(t)$ appearing
in Eqs. (\ref{imf1},\ref{imf2}), is defined through the matrix elements
\begin{eqnarray}
\bra{0}\bar{s}\gamma_\mu s\ket{K^-(k_1) K^+(k_2)}&=&(k_1-k_2)_\mu \fks(t)\\
\bra{0}\bar{s}\gamma_\mu s\ket{\bar{K}^0(k_1) K^0(k_2)}&=&(k_1-k_2)_\mu 
\fks(t)
\end{eqnarray}
with $\fks(0)=-1$.

\subsection{Kaon Strangeness Form Factor}
The appearance of the kaon's strangeness form factor, $\fks$, in
the expressions (\ref{imf1},\ref{imf2}) necessarily implies the 
introduction of some model
dependence into the dispersion relation analysis. The reason is that
there exists no data on $\fks(t)$. Consequently, the best we can do is 
illustrate the impact of choosing a reasonable parametrization of
this form factor. To this end, we first make a
few general observations regarding $\fks$ and its relationship to the
$K\bar{K}$ partial waves. In the product of $\fks(t)^{\ast}$ and the partial
waves $\bll$ appearing in Eqs. (\ref{imf1},\ref{imf2}), 
the real part will depend on both
the magnitudes of these two factors as well as on their relative phase.
Specifically, defining the phases as
\begin{eqnarray}
\label{p_b}
\bll&=&|\bll|\hbox{e}^{i\delta_1}\\
\label{p_fk}
\fks&=&|\fks|\hbox{e}^{i\delta_K}
\end{eqnarray}
one has
\begin{equation}
\label{rel_p}
\hbox{Re}\left\{\bll\fks(t)^{\ast}\right\}
=|\bll| |\fks| \cos(\delta_1-\delta_K)=|\bll| |\fks| (1+\gamma_K)
\end{equation}
where we define a phase difference correction $\gamma_K\equiv\cos(\delta_1-
\delta_K)-1$. 

The lack of data on $\fks$ is particularly problematic in
seeking to determine $\gamma_K$. Here, the situation stands in contrast to
the case of two pion contributions to the nucleon's isovector EM form
factors \cite{Hoe76,Mer96}. 
In the latter instance, the phase of the $\pi\pi$ partial
wave must be identical to that of the pion's isovector EM form factor for
$4 m_\pi^2\leq t\leq 16 m_\pi^2$. This feature follows from the fact that
in this kinematic range, there is only one final state (involving two
$\pi$'s) having the same quantum numbers as the isovector EM current. 
Unitarity then implies that  the phase of the form factor
and that of the scattering amplitude must be identical, 
that is, that the phase difference correction $\gamma_\pi=0$ 
\cite{Fub58}. In dispersion relation analyses
of the isovector form factors one typically assumes that $\gamma_\pi=0$
everywhere below the $N\bar{N}$ production threshold, since the phases
associated with $4\pi$, $6\pi$, {\em etc.} final states are small
\cite{Bas74}. This latter
practice falls under the rubric of \lq\lq extended unitarity" 
\cite{DrZ60,Fra59,Hoe75}.
In the case of $K\bar{K}$ scattering, however, there exist several 
multi-pion final states which can be reached for 
$t\geq 4m_K^2$. Hence, there exists
no regime in $t$ for which $\gamma_K=0$. At this time, we are unable
to make any statements about $\gamma_K$, and we take its value to be one
of the uncertainties in our analysis. We note, however, that $|1+\gamma_K|
\leq 1$. Thus, for purposes of setting an upper bound 
on the magnitude of the spectral function,
we may set $\gamma_K=0$.

In choosing our model parametrizations of $\fks(t)$ we draw upon what
is known about the lightest pseudoscalar meson form factors in the
time-like region. First, it is well-known that the pion's EM form
factor $\fpi(t)$ is dominated by the $\rho$-resonance for $4\mpis\leq t\leq
(\mpi+\mom)^2$ \cite{Hey81}. 
Moreover, more than 90\% of the pion charge radius can
be accounted for by the presence of a $\rho$-pole \cite{Eck89}. 
The simplest parametrization
which reproduces these gross features is that of the vector dominance
model (VDM). The detailed structure of $\fpi(t)$, including
the shape of the $\rho$-peak, requires more sophisticated 
parametrizations than that of $\rho$-dominance \cite{Hey81}. 
Nevertheless, one is able to approximate
the results of such analyses in the $\rho$-region using a 
VDM parametrization with values for $\mro$ and $\Gamma_\rho$ 
in good agreement with those
obtained from other observables \cite{Hey81,PDG96}. 
In the case of the kaon's EM form factor $\fk(t)$, one has information 
in the time-like region from $\sigma(e^+e^-\to K\bar{K})$ data 
\cite{Del81}. As extracted
from this data, $\fk(t)$ displays a peak near the $K\bar{K}$ threshold,
which is also close to the value $t=m_\phi^2$. Conventional treatments
of $\fk(t)$ have correspondingly employed extended versions of VDM,
including poles associated with not only the $\phi(1020)$ but also
the $\rho$ and $\omega$ \cite{Del81}. 
For values of $t\geq 2\ (\hbox{GeV}/c)^2$,
one begins to observe a bump-dip structure which cannot be reproduced
using the three lightest vector mesons, and one is apparently forced to
include poles associated with higher mass vector mesons 
\cite{Del81,Fel81}. 

For our present purpose, it is sufficient to choose a parametrization
for $\fks(t)$ which produces behavior in the time-like region in reasonable
accord with the gross structures of the pseudscalar EM form factors. 
Indeed, we are not interested in obtaining airtight numerical predictions 
for the nucleon's strangeness form factors, but rather in illustrating the
impact which the use of a realistic $\fks(t)$ has on these predictions. 
Hence, choosing a parametrization which produces correct structure in 
detail is not necessary. Because the current $\bar{s}\gamma_\mu s$ is
purely isoscalar, we expect no significant contribution from 
$1^+(1^{--})$ mesons such as the $\rho$.\footnote[1]{In short, we neglect 
isospin-breaking effects, such as $\rho$-$\omega$ mixing.} 
The lightest $0^-(1^{--})$ meson which might contribute is
the $\omega$. However, we would expect the matrix element $\bra{\omega}
\bar{s}\gamma_\mu s\ket{0}$ to be small since the $\omega$ is nearly
a pure $(\ket{u\bar{u}}+\ket{d\bar{d}})/\sqrt{2}$ state having a small
admixture of $\ket{s\bar{s}}$ at the level of $\epsilon\approx 0.05$. 
Consequently, we employ models which (a) are normalized
to give the correct strangeness charge, $\fks(0)=-1$ and (b) contain a
strong resonance enhancement in the vicinity of the $\phi(1020)$. 
The simplest such model is that of $\phi$-meson dominance, which yields
\begin{equation}
\label{fk_vdm}
|\fks(t)_{VDM}|=\left\{ {(\xi^2)^2+m_\phi^2\Gamma^2\over [(\xi^2-t)^2
+m_\phi^2\Gamma^2]}\right\}^{1/2}
\end{equation}
where $\xi^2\equiv m_\phi^2-\Gamma^2 /4$ and $\Gamma$ is the width of the
$\phi(1020)$ resonance. An alternative is to adopt the 
Gounaris-Sakurai (GS) parametrization, which is reasonably successful
in modelling $\fpi(t)$ in the $\rho$-peak region. When employing the
GS form, we replace the $\rho$ 
mass and width with those of the $\phi$. This parametrization can 
be found in Ref. \cite{GSa68} and we do not re-produce it here. 
It is interesting nevertheless to compare the VDM and
GS forms near the $\phi$-pole. Both can be shown to yield
\begin{equation}
\label{comp}
|\fks(t=m_\phi^2)|={m_\phi\over\Gamma}+\delta
\end{equation}
where $m_\phi/\Gamma\approx 255$, $\delta_{VDM}\leq 0.01$ and $\delta_{GS}
\approx -38$. We also note that both models fall off to unity from
their peak values at roughly the same place as $\fk(t)$ ($ t\approx
2\ (\hbox{GeV}/c)^2$. In the following discussion, we compare
predictions for the $\FIS(t)$ using the VDM and GS 
parametrizations with those obtained
assuming pointlike behavior, $\fks(t)\equiv -1$.

\section{Born Approximation and Beyond}
\label{sec:born}

Thus far, all calculations of the \lq\lq kaon cloud" continuum
contribution have been restricted to one loop order. In the case of
the non-linear SU(3) $\sigma$-model, for example, the relevant diagrams are
shown in Fig. 2. Performing such a one-loop calculation is equivalent
to (a) computing the amplitudes $\bra{N}\bar{J_N}\ket{n}V(\pbar)$
and $\bra{n} \bar{s}\gamma_\mu s\ket{0}$ entering the expression
in Eq. (\ref{spec_t}) under specific approximations and (b) using the
resultant spectral functions in the appropriate dispersion integral
of Eqs. (\ref{unsub}, \ref{sub}). 
In particular, for loop contributions where the current
is inserted on the kaon line (Fig. 2a), these approximations 
amount to computing
the $\bll$ in the Born approximation (see Fig. 3a) and taking the kaon's 
strange form factor to be point-like: $\fks\equiv -1$ (see Fig. 4a). 
For diagrams where the current is inserted on the strange baryon line 
(Fig. 2b), the corresponding approximations entail 
evaluating the $B\bar{B}\to N\bar{N}$ amplitude in the one meson exchange
approximation (Fig. 3b) and taking the strange baryon's strangeness form 
factor to be unity (Fig. 4b).  The remaining one-loop diagrams appearing
in Fig. 2c are needed to guarantee that the one-loop amplitudes satisfy
the Ward-Takahashi identity and have no analog within the framework of DR.
This equivalence between loops and DR 
has been discussed previously for the pion loop
contribution the the nucleon's isovector EM form factors in the context of 
the linear SU(2) $\sigma$-model \cite{Fed58,Che58,DrZ60}. 
In what follows, we demonstrate
the equivalence for the strangeness form factors using 
the non-linear SU(3) $\sigma$-model \cite{Geo84,com1}. 
We choose this model as it constitutes the standard
paradigm of a chiral effective theory.
We also show how, for the $K\bar{K}$ contribution
(Figs. 2a and 3a) the one-loop approximation is a rather drastic one.

In order to proceed, we first compute the $\bll$ in the Born approximation 
(B.A.), using the amplitudes associated with
the diagrams in Fig. \ref{Fig3}a. 
In the case of the baryon pole diagrams, we include
only the $\Lambda$ intermediate state since, in the limit of good SU(3)
symmetry, the strong $N\Sigma K$ coupling is highly suppressed with
respect to the $N\Lambda K$ coupling \cite{Hir71}. We obtain
\begin{eqnarray}
\label{bp_b}
\bpp&=& {1\over 12 \pi f^2}
\left({2\mn Q^2\over\sqrt{t}}\right)
\Biggl\{ {3\over 2}-{1\over 6}(3F+D)^2 \\
&&+{1\over 3}(3F+D)^2\left({{\bar M}^2 \over PQ}\right)[Q_0(\xbb-i\epsilon)
-Q_2(\xbb-i\epsilon)]\nonumber \\
&&+(3F+D)^2\left({{\bar M}^2 \over PQ}\right) Q_2
(\xbb-i\epsilon) \nonumber \\
&&+(3F+D)^2\left({{\bar M}^2 \Delta{\tilde M} \over Q^2}\right)
Q_1 (\xbb-i\epsilon)\Biggr\} \nonumber \\
\label{bm_b}
\bpm&=&{1\over 12 \pi f^2 }\left({2\sqrt{2}EQ^2
\over\sqrt{t}}\right)
\Biggl\{ {3\over 2}-{1\over 6}(3F+D)^2 \\
&&+{1 \over 3}(3F+D)^2\left({{\bar M}^2 \over PQ}\right)[Q_0(\xbb-i\epsilon)
-Q_2(\xbb-i\epsilon)]\Biggr\} \nonumber
\end{eqnarray}
where
\begin{eqnarray}
\xbb&=&\nu_p/\nu_0 \nonumber \\
\nu_p&=&\nu_B+{\bar M}\Delta{\tilde M} \nonumber \\
\nu_B&=&(t-2 m_K^2)/4\mn \label{def_b} \\
\nu_0&=&PQ/\mn \nonumber \\
{\bar M}&=&(\mn+m_\Lambda)/2 \nonumber \\
\Delta {\tilde M}&=&(m_\Lambda-\mn)/\mn \nonumber
\end{eqnarray}
where $f \approx 93 \mbox{ MeV}$ is the pion decay constant,
and where the $Q_n(z)$ are Legendre functions of the second kind.
The constants $F$ and $D$ are just
the usual SU(3) reduced matrix elements, with $D+F=1.26$ and $F/D=0.64$.
Substituting these expressions into
the formulae of Eqs. (\ref{imf1}) and (\ref{imf2}) yields
\begin{eqnarray}
\label{foba}
\hbox{Im}\  \FOS(t) &=&
{ 1 \over 12 \pi f^2} \left({Q^3 \over 2 \sqrt{t}}\right)
\hbox{Re}\left[\fks(t)\right]
\Biggl\{ {3\over 2}-{1\over 6}(3F+D)^2\\
&&+{1\over 3}(3F+D)^2\left({{\bar M}^2 \over P Q}\right)
\left[Q_0(\xbb)-Q_2(\xbb)\right] \nonumber \\
&&-(3F+D)^2 {\mns \over P^2}\left[
\left({ {\bar M}^2 \over P Q}\right) Q_2(\xbb)
+\left({{\bar M}^2 \Delta{\tilde M} \over Q^2}
\right) Q_1(\xbb) \right]\Biggr\} \nonumber \\
\label{ftba}
\hbox{Im}\  \FTS(t) &=&
{ 1 \over 12 \pi f^2} \left({Q^3 \over 2 \sqrt{t}}\right)
\hbox{Re}\left[\fks(t)\right]\;
(3F+D)^2 {\mns \over P^2}\\
&&\left\{ \left({{\bar M}^2 \over P Q}\right) 
Q_2(\xbb) +\left({{\bar M}^2 \Delta{\tilde M} \over Q^2}
\right) Q_1(\xbb)\right\} \nonumber \ \  ,
\end{eqnarray}
where we have made use of the fact that the $\bll$ are 
real in the B.A. for $t \geq 4m_K^2$.

After setting $\fks(t)\equiv -1$ in Eqs. (\ref{foba}) and 
(\ref{ftba}), one obtains expressions for the spectral functions which
are identical to those obtained from the Feynman amplitudes associated
with the diagrams in Fig. 2a. To see how this equivalance comes about, we
refer to the analytic structure of the matrix element
$\bra{N(p); N(\pbar)}\bar{s}\gamma_\mu s\ket{0}$. Any discontinuities
across the real $t$-axis
must arise from integration over poles associated with the presence of
one of the physical states $\ket{n}$ appearing in Eq. (\ref{spec_t}). 
The Cutkosky rules \cite{Cut60,IZu80} give a procedure for extracting
these discontinuities from Feynman Amplitudes. 
In particular, we may obtain the corresponding
discontintuity from the Feynman amplitudes by making the following 
replacement for each propagator associated with one of the particles
appearing in the given state $\ket{n}$:
\begin{equation}
\label{cutko}
{1\over p^2-m^2+i\epsilon}\longrightarrow -2\pi i\, \theta(p_0)
\,\delta(p^2-m^2) \ \ \ .
\end{equation}
Since the only state $\ket{n}$ contained in the loops of
Fig. 2a is $\ket{K\bar{K}}$, we make the replacement of Eq. (\ref{cutko})
for the two kaon propagators in the loop integrals. Doing so, and carrying
out the loop integration, yields the formulae in Eqs. 
(\ref{foba}, \ref{ftba}).
The details of this procedure are shown in the Appendix. Thus, insofar 
as the DR of Eqs. (\ref{unsub}, \ref{sub}) are valid, the use of 
one-loop amplitudes and the use of Eq. (\ref{spec_t}) with the B.A. 
for the $K\bar{K}\to N\bar{N}$ amplitudes are equivalent. 

With explicit formulae for the spectral functions in hand, 
it is now straightforward to carry out the dispersion integrals.
When the non-linear SU(3) $\sigma$-model is used to 
perform one-loop calculations for these leading moments, one finds that 
$\rho^s_D$ contains a U.V. divergence. Using the dispersion relation
framework, we correspondingly find that the $K\bar{K}$ contribution
to $\rho^s_D$ is divergent in the dispersive variable $t$ when the B.A.
is used to compute the $\bll$ and a pointlike kaon strangeness form factor
is employed. In the case of loops, this U.V. divergence can
be handled in a variety of ways. When one attempts an analysis using
CHPT, the divergence is removed by the corresponding counterterm. This 
counterterm, however, contains a finite remainder which cannot be determined 
in any model-independent way from existing measurements \cite{Mus96}. 
Consequently, one must invoke additional, model-dependent assumptions in 
order to make predictions using loops. A variety of such scenarios are 
discussed and evaluated in  Ref. \cite{Mus96}. These alternatives include 
assuming the finite low-energy constants in CHPT are saturated by vector 
meson resonances or assuming that the loop integrals are cut-off by form 
factors and the meson-baryon vertices. Each involves a departure from QCD 
(at the level of hadronic effective theory) to a greater or lesser extent 
and entails a certain amount of ambiguity. Ideally, one would like to 
find a less model-dependent way of regulating the U.V. behavior of the 
integrals and obtaining a finite prediction.

In the present context, the unitarity bound on the partial waves (Eqs. 
(\ref{par_p},\ref{par_m}))
provides such a model-independent regulator. The physical amplitudes
$\bll$ must satisfy the bound (Eq. (\ref{ubs})), 
regardless of one's model for $KN$
scattering. To illustrate the impact of the unitarity bound, we plot in
Fig. \ref{Fig5}  the partial waves computed in the B.A. as a function
of $t$ and the corresponding unitarity bound above the two nucleon threshold. 
One sees that the $\bll$ in the B.A. violate the unitarity
bound by a factor of four or more at threshold, and that this violation
grows with $t$. 

When translating the unitarity bound into a bound on the spectral functions,
some care is required. The most na\"\i ve approach is to begin with
Eqs. (\ref{imf1}) and (\ref{imf2}), apply the triangle inequality,
and take $|\bll|=1$, {\em viz}
\begin{eqnarray}
\label{fobn}
|\hbox{Im}\  \FOS(t)| & \leq & \left({\mn Q\over 4 P^2}\right)\bigg\vert
{E\over\sqrt{2}\mn}\bpm-\bpp\bigg\vert \bigg|\fks(t)^{\ast}\bigg| \\
& \leq & \left({\mn Q\over 4 P^2}\right)\left\{{E\over\sqrt{2}\mn}
\bigg|\bpm\bigg|
+\bigg|\bpp\bigg|\right\}\bigg|\fks(t)\bigg|\ \ \ , \nonumber
\end{eqnarray}
and similarly for $|\hbox{Im}\ \FTS(t)|$.
In arriving at the first line of Eq. (\ref{fobn}) we have set
the phase difference correction $\gamma_K=0$ as discussed previously.
Setting $|\bll|=1$ and using $t=\sqrt{2}E$ in the $N\bar{N}$ c.m. frame, 
we obtain the
na\"\i ve unitarity bounds
\begin{eqnarray}
\label{ubf1n}
|\hbox{Im}\ \FOS(t)| &\leq& \frac{Q}{8\sqrt{2} P^2}
\left( 2\sqrt{2}\mn + \sqrt{t} \right)|\fks(t)|\, ,\\
\label{ubf2n}
|\hbox{Im}\ \FTS(t)| &\leq& \frac{\mn Q}{4\sqrt{2 t} P^2}
\left( \sqrt{2 t} + 2 \mn\right)|\fks(t)|.
\end{eqnarray}
These na\"\i ve  bounds (Eqs. (\ref{ubf1n},\ref{ubf2n})) are shown in 
Fig. \ref{Fig6} together with the the B.A. where a pointlike strangeness 
form factor for the kaon has been applied. The divergence in these bounds
appearing at the $N\bar{N}$ threshold arises from the $1/P^2$ factor 
appearing
in Eqs. (\ref{ubf1n}) and (\ref{ubf2n}). The presence of this singularity
renders the functions appearing in the RHS of Eqs. (\ref{ubf1n}) and 
(\ref{ubf2n}) non-integrable over the range $4\mns\leq t\leq \infty$.
Thus, the na\"\i ve bounds are not meaningful.

A more careful application of unitary requires that one also take into
account the threshold relation on the $\bll$ appearing in Eq. (\ref{th_rel}).
This relation forces the linear combinations of $\bll$ 
appearing in Eqs. (\ref{imf1}) and (\ref{imf2})
to go as $P^2$ near threshold, thereby ensuring that the spectral functions 
are finite as $P\to 0$. Hence, when imposing unitarity, one must enforce
the threshold relation. For simplicity, we choose to take
$\bpp=\bpm/\sqrt{2}$ everywhere above $4\mns$, even though this relation
rigorously applies only at $t=4\mns$, and take $|\bpm|\leq 1$. 
This leads to the bounds
\begin{eqnarray}
\label{ubf1th}
|\hbox{Im}\ \FOS(t)| &\leq& \frac{Q}{2\sqrt{2} ( \sqrt{t} +2 \mn)}|\fks(t)|
\, ,\\
\label{ubf2th}
|\hbox{Im}\ \FTS(t)| &\leq& \frac{ \mn Q}{\sqrt{2 t} (\sqrt{t} +2 \mn)}
|\fks(t)| \, ,
\end{eqnarray}
which now can be used in the dispersion relations Eqs. (\ref{unsub},
\ref{sub}) without ambiguity.
Furthermore, the bounds with the correct threshold behaviour built in 
are always more stringent than the na\"\i ve ones for all $t \geq 0$.
Fig. \ref{Fig7} shows these bounds (Eqs. (\ref{ubf1th},\ref{ubf2th}))
together with the B.A. and a pointlike kaon strangeness form 
factor in both cases. We show only the bound on 
$|\hbox{Im}\ \FOS(t)|$, since $|\hbox{Im}\ \FTS(t)|\leq 
|\hbox{Im}\ \FOS(t)|\,m/E$.
It is clear from the curves in Fig. \ref{Fig7} that unitarity has a
significant impact on the spectral functions above the $N\bar{N}$ threshold.

In addition to correcting the $K\bar{K}\to N\bar{N}$ in the B.A. 
for unitarity,
we also attempt a more realistic treatment of the kaon's 
strangeness form factor appearing in Eqs. (\ref{imf1}) and (\ref{imf2}).
As discussed above, we do so by choosing two parametrizations strongly 
peaked in the vicinity of
the $\phi(1020)$ resonance. In Fig. \ref{Fig8} we plot the same quantities
as in Fig. \ref{Fig7} but using the GS form factor. For $4 m_K^2\leq t
\leq 4\mns$, the $\phi$ peak in the GS parametrization leads to a strong 
enhancement of the spectral functions as compared with the use of a 
pointlike
form factor. As $t$ increases beyond the $N\bar{N}$ threshold, the GS
form factor eventually suppresses the spectral functions when either
the B.A. or unitarity bounds are used. The impact of using the simpler
VDM parametrization is similar to that of the GS form factor. Although
we could have attempted to carry out a more detailed analysis of $\fks(t)$,
the plot in Fig. \ref{Fig8} makes the essential point clear: the impact of
choosing a reasonable, non-pointlike form for $\fks(t)$ can be non-trivial.

\section{Strangeness Moments}
\label{results}

In this section, we explore the numerical consequences of unitarity
and $\fks(t)$ parametrization for the leading strangeness moments,
$\rho^s_D$ and $\mu^s$. For purposes of later discussion, it is useful
to write down the DR for these two quantities:
\begin{eqnarray}
\label{rhosi}
\rho^s_D&=&-{4\mns\over\pi}\int_{4 m_K^2}^\infty dt {\hbox{Im}\ \FOS(t)
	\over t^2} \\
\label{musi}
\mu^s&=& {1\over\pi} \int_{4 m_K^2}^\infty dt {\hbox{Im}\ \FTS(t)\over t}
\ \ \ .
\end{eqnarray}

Using these expressions, we compare three scenarios for
computing the $K\bar{K}$ contribution to the moments: (a) a calculation
using the B.A. for the $\bll$ and pointlike kaon strangeness form
factor (BA/PFF); (b) the same as (a) but imposing the unitarity bounds
of Eqs. (\ref{ubf1th}) and (\ref{ubf2th}) for $t\geq 4\mns$ (BA/U/PFF); 
(c) the same as (b) but using the GS parametrization for $\fks(t)$  
(BA/U/GS). Of these scenarios, we recall that (a) is equivalent to computing
the one-loop amplitudes of Fig. 2a. We further delineate between the 
contributions to the dispersion integrals in Eqs. (\ref{sub}) and
(\ref{unsub}) arising from the integration regions $4 m_K^2\leq t\leq 4\mns$
and $4\mns\leq t$. In applying the unitarity bound (scenarios (b) and 
(c)), we assume for simplicity that the spectral functions do not change
sign across the two nucleon threshold and that this sign is given by the
phase of the spectral function for $t\leq 4\mns$.
The results are given in Table I.

$$\hbox{\vbox{\offinterlineskip
\def\strut{\hbox{\vrule height 15pt depth 10pt width 0pt}}
\hrule
\halign{
\strut\vrule#\tabskip 0.2cm&
\hfil$#$\hfil&
\vrule#&
\hfil$#$\hfil&
\vrule#&
\hfil$#$\hfil&
\vrule#&
\hfil$#$\hfil&
\vrule#&
\hfil$#$\hfil&
\vrule#\tabskip 0.0in\cr
& \multispan9{\hfil\bf TABLE I\hfil} & \cr\noalign{\hrule}
& \hbox{Moment} && \hbox{Scenario} && 4 m_K^2\leq t\leq 4\mns && 4\mns\leq t
&& \hbox{Total}
& \cr\noalign{\hrule}
& \rho^s_D&& \hbox{BA/PFF} && 0.18 && \hbox{div} &&\hbox{div} & \cr
& && \hbox{BA/U/PFF} && 0.18 && 0.03 && 0.21 &\cr
& && \hbox{BA/U/GS} && 0.26 && 0.01 && 0.27 &\cr
\noalign{\hrule}
& \mu^s && \hbox{BA/PFF} && -0.07 && -0.40 && -0.47 &\cr
& && \hbox{BA/U/PFF} && -0.07 && -0.07 && -0.14 &\cr
& && \hbox{BA/U/GS} && -0.09 && -0.01 && -0.10 & \cr
\noalign{\hrule}}}}$$
{\noindent\narrower {\bf Table I.} \quad Contributions from kaon
intermediate state to the nucleon's strangeness radius and magnetic moment,
computed using dispersion relations. Results are given using three
different scenarios as discussed in the text: (a) BA/PFF: partial waves
$\bll$ computed in B.A. and kaon strangeness form factor $\fks(t)\equiv 
-1$, (b) BA/U/PFF: same as (a) but with unitarity limit from Eqs.
(\protect\ref{ubf1th},\protect\ref{ubf2th}) applied for $t\geq 4\mns$,
(c) BA/U/GS: same as (b) but with Gounaris-Sakurai parametrization
for $\fks(t)$. To convert $\rho^s_D$
to $\langle r_s^2\rangle$, multiply $\rho^s_D$ by $-0.066$ fm$^2$.
\smallskip}

From the entries in the table, the numerical impact of imposing
unitarity and choosing a non-pointlike form factor is evident. In the
case of $\rho^s_D$, unitarity eliminates the U.V. divergence and sets
a bound on the contribution from the region above the $N\bar{N}$ threshold
which is small. In terms of the dimensionfull Dirac radius, this 
contribution is about $-0.002$ fm$^2$. The use of the GS parametrization
for $\fks(t)$, on the other hand, increases the contribution from the
region $4 m_K^2\leq t\leq 4\mns$ by about 50\%, owing largely to the
$\phi$  peak near the two kaon threshold. Even though the $F_1$ spectral
function with the GS form factor falls below the corresponding spectral
function with a pointlike form factor for $t > 2\ (\hbox{GeV}/c)^2$, the
$1/t^2$ appearing in the integrand of Eq. (\ref{rhosi}) favors the
contribution from the region containing the $\phi$-resonance enhancement.
Consequently, the reduction for $2\ (\hbox{GeV}/c)^2\leq t\leq 4\mns$
is not significant.

For the strange magnetic moment, the B.A. contribution with a pointlike
$\fks(t)$ yields a finite result, in contrast to the situation with
$\rho^s_D$. Nevertheless, the imposition of unitarity reduces the
$t\geq 4\mns$ contribution to one sixth of its B.A. value. Insofar as
the contribution from this region was the dominant one in the B.A., 
this unitarity reduction is quite significant. The use of the GS form
factor reduces this contribution even further, whereas its impact in
the region $4 m_K^2\leq t\leq 4\mns$ is small. In the latter instance,
the enhancement from the $\phi$ peak is not as important as in the case of
$\rho^s_D$, since the integrand in Eq. (\ref{musi})
only weights the low-$t$ behavior as $1/t$. 

We emphasize that, although the results listed in the last column of Table
I may be instructive, one should not take the precise numerical values
too seriously. It is clear
from the results in the fourth column -- as well as from the curves in Figs.
\ref{Fig5}, \ref{Fig7}, and \ref{Fig8} -- that the consequences of the
unitarity constraints are significant. The physical mechanisms responsible
for the reduction of the $\bll$ and $\hbox{Im}\  F_i^s$ from their B.A. 
values
to the unitarity limits -- primarily non-resonant and resonant kaon
rescattering -- cannot be neglected in a physically realistic calculation.
Although the unitarity bounds give an explicit indication of the importance
of these rescattering terms in the region $t\geq 4\mns$, one has no reason
to assume they are any less important in the region $4 m_K^2\leq t\leq
4\mns$. Whether rescattering effects increase or decrease the contribution
from this region is not known at present, and one may only speculate. For
example, the presence of a $\phi(1020)$ resonance in the $K\bar{K}\to
N\bar{N}$ partial waves could, in principle, enhance the $\bll$ from
their B.A. values in some region of $t$. In fact, previous experience
with $\pi\pi$ contributions to
nucleon's isovector form factors suggests that rescattering may
lead to enhanced low-$t$ contributions. In the work of Ref. \cite{Fed58},
it was found that, in comparison to the B.A. contribution,
rescattering contributions enhanced the $t\leq 4\mns$ contribution
to the isovector magnetic moment by roughly the same
magnitude as the unitarity bounds reduced the $t\geq 4\mns$ contribution.

Given the equivalence between the BA/PFF treatment of the dispersion
relation and the one loop contribution of Fig. 2a, the results of
the foregoing analysis should lead one to question the credibility
of any one-loop prediction for the strangeness moments. Even model 
calculations
which employ form factors to regulate the integrals do not include
all of the rescattering corrections required by unitarity. Indeed, such
form factors apply only the to the meson-nucleon vertices, and not to the
full $K\bar{K}\to N\bar{N}$ (or $KN\to KN$) scattering amplitude. Moreover,
meson-nucleon form factors are often taken to be functions of $k^2$,
where $k_\mu$ is the four-momentum of the kaon, and are normalized to
reproduce the SU(3) values for the meson-nucleon
coupling when $k^2=m_K^2$. Thus, hadronic form factors have no impact on 
the B.A. violation of unitarity for scattering amplitudes in the physical 
region. 

In a similar vein, we note that the use of a point-like kaon strangeness
form factor, as is used in most loop calculations reported to date,
could represent as serious an error as the violation of unitarity
in the B.A. $\bll$. A comparison of the BA/U/PFF and BA/U/GS results
in Table I shows that the inclusion of a reasonable parametrization of
$\fks(t)$, displaying an enhancement in the vicinity of the $\phi(1020)$
can change magnitudes of $\rho^s_D$ and $\mu^s$ by as much as 30\%. While
our rationale for choosing such a parametrization is not based on any
rigorous argument, we nevertheless believe that it constitutes a more
realistic input than does the use of a pointlike form factor. We
correspondingly expect most one-loop calculations employing the pointlike
approximation to be physically un-realistic. 

As a final observation, we make a comparison between the DR calculation
and the one kaon-loop calculation of CHPT. To be concrete, we focus
on the strangeness radius. Within the framework of CHPT, the only
well-defined piece of a one kaon loop contribution to $\rho^s_D$ is that 
which is non-analytic in the strange quark mass. The remaining piece is
indistinguishable from tree-level contributions arising from the chiral
Lagrangian, at a given order in the chiral scale, $\lamchi\approx 4\pi f$. 
Consequently,
one subsumes all analytic contributions into the counterterms. In the
case of $\rho^s_D$, only the amplitudes of Fig. 2a contribute a term
non-analytic in $m_s$ at ${\cal O}(1/\Lambda_\chi^2)$. Specificially,
one finds \cite{Mus96}
\begin{equation}
\label{chpta}
\rho^s_D=\rho^s_\sst{LOOP}-\left({2\mn\over\lamchi}\right)^2 c^s\ \ \ ,
\end{equation}
where
\begin{equation}
\label{chptb}
\rho^s_\sst{LOOP}=\left({\mn\over\lamchi}\right)^2\left\{1+{5\over 3}\left[
	\left({3F+D\over\sqrt{6}}\right)^2+{3\over 2}(D-F)^2\right]\right\}
	\left[{\cal C}_\infty-\ln{m_K^2\over\mu^2}\right] \ \ \ ,
\end{equation}
where ${\cal C}_\infty$ contains the U.V. regulator and $\mu$ is the
renormalization scale.\footnote{In Ref. \protect\cite{Mus96} contributions
from $\Sigma K$ intermediate states were also included, yielding the
term proportional to $(D-F)^2$ in Eq. (\protect\ref{chptb}). This
contribution has been omitted in the present analysis.} The counterterm
$c^s$ contains a piece cancelling the U.V. divergence appearing in 
$\rho^s_\sst{LOOP}$ plus a finite remainder, containing all the analytic
contributions at order $1/\Lambda_\chi^2$. The finite part of $c^s$ can
be further decomposed as
\begin{equation}
c^s=c_0-2[c_{-}-(c_{+}/3)]\ \ \ ,
\end{equation} 
where the the $c_{\pm}$ can be determined from the neutron and proton
EM charge radii and where the constant $c_0$ is associated with the 
SU(3) singlet current. It is the latter constant which cannot be determined
from any existing data, since measurements have only been made of SU(3)
octet vector current matrix elements. Consequently, CHPT cannot be used to 
make a model-independent prediction for $\rho^s_D$. 

The correspondence between the results in Eqs. (\ref{chpta}, \ref{chptb})
and those obtained using the dispersion relation 
can be understood as follows. In the B.A. with a pointlike kaon form factor,
one finds an identical $\ln m_K^2$ I.R. singularity as that appearing in 
$\rho^s_\sst{LOOP}$. The origin of this $\ln m_K^2$ is a branch cut
singularity in the B.A. partial waves for $t\leq 4 m_K^2(1- m_K^2/4\mns)$
\cite{Mer96,DrZ60}. 
The dispersion relation result contains no 
renormalization scale dependence since the
unitarity bound removes the U.V. divergence. As the nucleon mass is
the only other scale which enters the calculation, one finds $\mu\to
\mn$ in the leading logarithmic contribution. 
Presumably, the remaining contributions
in the B.A., as well as those generated by resonant and non-resonant
rescattering terms in the $\bll$ (as necessitated by unitarity) and
the effects of the physical $\fks(t)$, should be accounted for in 
CHPT by the counterterm, $c^s$. Unfortunately, since $c^s$ cannot
be determined from existing data using symmetry, one must resort to
other strategies for including the rescattering and $\fks(t)$ effects.

\section{Summary}
\label{summary}

In the present paper, we have made an initial study of the continuum
contribution to the nucleon's strangeness vector current form factors
using the framework of dispersion relations. In focusing on the
$K\bar{K}$ contribution, we have illustrated how a leading order
loop prediction for the strangeness radius and magnetic moment entails
a substantial violation of unitarity. At the same time, we have derived
a unitarity bound on this continuum contribution from the region in
the dispersion integral above the $N\bar{N}$ production threshold. 
Although we have specified our analysis to the case of the non-linear 
SU(3) $\sigma$-model, our conclusions regarding unitarity violation
should hold for any chiral model which yields a similar
structure for the $K\bar{K}\to N\bar{N}$ scattering amplitude in the 
Born approximation. Our statement of the unitarity bound is general.
We have also illustrated how the use of a reasonable, realistic kaon
strangeness form factor can significantly affect one's predictions for
$\rho^s$ and $\mu^s$. We conclude that most model predictions for the
two kaon continuum contributions are physically un-realistic. We further
suspect that our conclusions regarding the $K\bar{K}$
intermediate state ought to apply as well to other leading-order
loop calculations, whether they involve higher-mass strange mesons and
baryons -- as in the quark model calculation of Ref. \cite{Gei96} -- 
or states containing three or more pseudoscalar mesons. 

We emphasize that the contribution about which we have yet to 
make a definitive 
statement is the $K\bar{K}$ contribution from the region below the
$N\bar{N}$ threshold. At present, the best we can do is make an estimate
based on the B.A. for the $\bll$ and a non-pointlike kaon strangeness form
factor. The feasibility of making a refined analysis of
this contribution by continuing fits to physical $KN\to KN$ or $K\bar{K}
\to N\bar{N}$ scattering data will be discussed in a forthcoming study.
Nevertheless, we are able to show how the contribution from
this region to $\rho^s_D$ can be significantly enhanced if the kaon's
strangeness form factor is strongly peaked in the vicinity of the
$\phi(1020)$, as one would reasonably expect based on analogy with
$e^{+}e^{-}\to K\bar{K}$ data and on the flavor content of the lowest-lying
$0^{-}(1^{--})$ mesons. What remains to be resolved is the discrepancy 
between predictions for $\rho^s_D$ using a VDM approach and those obtained
using models for the continuum. The key may lie in a better understanding
of the sub-threshold behaviour of the $\bll$ as well as of the 
the contribution from the three pion continuum. Although it
contains no valence strange quarks, the latter is the lighest state which may
contribute to the DR for $\FOS$ and $\FTS$. The scale of this contribution,
along with those of $N\bar{N}$ and $B\bar{B}$ intermediate states, awaits
the result of future work.

\acknowledgements
We wish to thank R.L. Jaffe, N. Isgur, and U.-G. Mei{\ss}ner for
useful discussions. HWH has been supported by the Deutsche
Forschungsgemeinschaft (SFB 201) and the German Academic Exchange Service
(Doktorandenstipendium HSP II/ AUFE). MJM has been supported
in part under U.S. Department of Energy contracts \# DE-FG06-90ER40561
and \# DE-AC05-84ER40150 and under a National Science Foundation
Young Investigator Award.


\renewcommand{\Re}{\mbox{\rm {Re}}}
\renewcommand{\Im}{\mbox{\rm {Im}}}
\newcommand{\dida}[1]{/ \!\!\! #1}
\newcommand{\didag}[1]{/ \!\!\!\!\!\, #1}
\newcommand{\nomi}{\hphantom{-}}

\appendix
\section{Imaginary parts to one loop}
We show here the equivalence between the one-loop diagrams
of Fig. \ref{Fig2}a and the Born approximation for the $KN$-scattering
amplitudes in conjunction with the dispersion relation approach
( See Fig. \ref{Fig3}a). To that end, we calculate the imaginary 
part of the one-loop diagrams from Fig. \ref{Fig2}a which arises from 
the t-channel discontinuity.
The equality is then easily checked by comparing our results
with Eqs. (\ref{foba},\ref{ftba}). It does not depend on
one's choice for the kaon strangeness form factor.
For simplicity, we therefore assume pointlike kaons. 
Any non-pointlike kaon strangeness
form factor would simply multiply the resulting spectral functions.

In the following, we refer to the diagram with the propagating
$\Lambda$ (the triangle diagram) as diagram (1).
We assign the momenta to the particle lines as
shown in Fig. 9. For the other diagram with the kaon loop
(refered as diagram (2)), we assign the momenta in the same
way and leave out the $\Lambda$-momentum.
Since we produce a nucleon-antinucleon pair,
$q$ has to be timelike, i.e. $q^2=t\geq 0$.
We work in the center-of-momentum frame
of the nucleon-antinucleon pair, where $q=(\omega,\vec{0})$.
Using momentum conservation, we have $p'=(\omega/2,\vec{p'})$ 
and $p=(\omega/2,-\vec{p'})$ with $|\vec{p'}|=P=\sqrt{t/4-m_N^2}$.
We define the contribution of a particular Feynman
diagram to the  vertex function $\Gamma_\mu$ by
\begin{equation}
  \label{ver}
{\cal M}_\mu^{(i)}= -i\, \bar{U}(p') \Gamma^{(i)}_\mu V(p)\; ,  
\end{equation}
where the strangeness charge of the 
kaons $Q_s \equiv -1$ has been absorbed in $\Gamma_\mu^{(i)}$. 
Using the nonlinear $SU(3)$ $\sigma$-model and calculating
the isoscalar contribution, we obtain the following
contributions to the vertex functions : 
\begin{eqnarray}
\Gamma_\mu^{(1)} &=& i Q_s\frac{(3F+D)^2}{6 f^2}
\int\frac{d^4 k}{(2\pi)^4} \label{v1} \\
& &\frac{k_\mu (\dida{k}+\dida{q}/2)(\dida{p}'-\dida{k}-\dida{q}/2
-m_\Lambda)(\dida{k}-\dida{q}/2)}{[(k-q/2)^2 -m_{K}^2+i\epsilon]
[(k+q/2)^2 -m_K^2 +i\epsilon]
[(p'-k-q/2)^2 -m_{\Lambda}^2]} \; \nonumber \\
\Gamma_\mu^{(2)} &=& i Q_s\frac{3}{f^2} \int\frac{d^4 k}{(2\pi)^4}
\frac{k_\mu\dida{k}}{
[(k-q/2)^2 -m_K^2 +i\epsilon][(k+q/2)^2 -m_K^2 +i\epsilon]} \; 
\label{v2}
\end{eqnarray}
Since the denominator of the $\Lambda$-propagator does 
not vanish in the t-channel physical region the $i\epsilon$
can be dropped. The $\Gamma^{(i)}_\mu$ have branch cuts on
the real axis for $ t \geq 4 m_K^2$.
We calculate now the imaginary parts stemming from
the discontinuity associated with these cuts,
\begin{equation}
  \label{discon}
\Im\,\Gamma^\mu = \frac{1}{2\,i}\Delta\Gamma^\mu =
\frac{1}{2\,i}\lim_{\delta \to 0}
\left(\Gamma^\mu(\omega+i\delta)-\Gamma^\mu(\omega-i\delta)\right) \, .  
\end{equation} 
It is convenient to use the so-called Cutkosky rules 
\cite{Cut60,IZu80}, which give a compact expression for
the discontinuities associated with physical region singularities
of Feynman amplitudes. In particular, 
we obtain the discontinuities $\Delta\Gamma^{(i)}_\mu$ by
cutting the kaon lines in the diagrams (1) and (2) and replacing 
their propagators by $\delta$-functions (\ref{cuma}),
\begin{equation}
  \label{cuma}
\frac{1}{p^2 - m^2 + i\epsilon} \longrightarrow
-2\pi i\,\theta(p_0)\, \delta(p^2 - m^2)\; .
\end{equation}
As a consequence, the discontinuity arises for the intermediate
particles on the mass-shell. Note the equivalence to the
dispersion relation approach, in which the intermediate states
are also on-shell.
Due to the $\delta$-functions, the $d^4 k$ integration now
covers only a finite part of the $k$ space, leading to a finite  
value of the integral. Consequently, the divergences of the
integrals Eqs. (\ref{v1},\ref{v2}) do not contribute to the
discontinuity across the cut.
The imaginary part is finite, only the real part has to be 
regulated.
Next we write $d^4 k$ as $dk_0\, k^2 dk \,d\Omega_k$ and use the 
$\delta$-functions to carry out the $dk_0$ and $dk$ integrations.
As a consequence, we obtain $k=(0,\vec{k})$ with
$|\vec{k}|=Q=\sqrt{t/4-m_K^2}$.
Moreover, the $d\Omega_k$ integration involves only
the cosine $x$ of the angle between $\vec{k}$ and $\vec{p'}$.
We obtain
\begin{eqnarray}
\Im\,\Gamma^{(1)}_\mu &=& Q_s\frac{(3F+D)^2}{48 \pi f^2}\frac{Q}{\sqrt{t}}
\frac{1}{2}\int_{-1}^1 dx \,\left( k_\mu(\dida{k}+2 {\bar M})
-\frac{2 {\bar M}^2}{P Q}\frac{k_\mu(\dida{k}+ 
m_N \Delta{\tilde M})}{\xbb -x}\right) \; , \\
\Im\,\Gamma^{(2)}_\mu &=& -Q_s\frac{3}{8 \pi f^2} \frac{Q}{
\sqrt{t}}\frac{1}{2}\int_{-1}^1 dx \, k_\mu \dida{k} \; ,
\end{eqnarray}
where
\begin{equation}
\xbb = \frac{t-2 m_K^2+4 m_N{\bar M}\Delta{\tilde M}}{4 P Q} > 1 \; .
\end{equation}
Finally,
$\Im\,\Gamma^{(i)}_\mu$ can be expressed in terms of the integrals
\begin{eqnarray}
  \label{k_ints}
L_\mu &=& \frac{1}{2}\int_{-1}^1 dx \, k_\mu\; ,\\
L_{\mu\nu} &=& \frac{1}{2}\int_{-1}^1 dx \, k_\mu k_\nu\; ,\\ 
I_\mu &=& \frac{1}{2}\int_{-1}^1 dx \, \frac{k_\mu}{\xbb - x} \; ,\\
I_{\mu\nu} &=& \frac{1}{2}\int_{-1}^1 dx \, \frac{k_\mu\,k_\nu}{\xbb - x}
\; , 
\end{eqnarray}
and these integrals can be decomposed into $g_{\mu\nu}$ and
symmetrical combinations of the independent four-vectors 
$\Delta=(p'-p)/2$ and $q=p+p'$. Their coefficients
can be obtained in a standard manner by evaluating the integrals
$q_\mu I^\mu, \Delta_\mu I^\mu$ and so on.
Furthermore, the I-integrals 
can be expressed through Legendre functions of the second kind.
For example, we find
\begin{eqnarray}
I_{\mu\nu}&=& \frac{1}{3} Q^2 \left[Q_2(\xbb)-Q_0(\xbb)\right] g_{\mu\nu}
-\frac{1}{3}\frac{Q^2}{t} \left[Q_2(\xbb)-Q_0(\xbb)\right] q_\mu q_\nu
\nonumber \\ \label{ex}
& &+\frac{Q^2}{P^2} Q_2(\xbb) \Delta_\mu \Delta_\nu \; .
\end{eqnarray}
Using the relation
\begin{equation}
  \label{f_zer}
\Im \,\Gamma^{(i)}_\mu = \gamma_\mu \Im\,F^{(i)}_1 
+i\frac{\sigma_{\mu\nu}}{2m}q^\nu \Im\, F^{(i)}_2 \; ,
\end{equation}
we can identify the contributions to the imaginary parts of the 
Dirac and Pauli form factors for $t \geq 4 m_N^2$, respectively.
We add now the contributions of the two diagrams and 
the spectral functions emerge as
\begin{eqnarray}
\label{specfunc1}
\hbox{Im}\  \FOS(t) &=&
{ Q_s \over 24 \pi f^2}\, {Q^3 \over \sqrt{t}}\,
\Biggl\{ {3\over 2}-{1\over 6}(3F+D)^2 \\
&&+{1\over 3}(3F+D)^2\left({{\bar M}^2 \over P Q}\right)
\left[Q_0(\xbb)-Q_2(\xbb)\right] \nonumber \\
&&-(3F+D)^2 {\mns \over P^2}\left[
\left({ {\bar M}^2 \over P Q}\right) 
Q_2(\xbb)
+\left({{\bar M}^2 \Delta{\tilde M} \over Q^2}
\right) Q_1(\xbb) \right]\Biggr\} \nonumber \\
\label{specfunc2}
\hbox{Im}\  \FTS(t) &=&
{ Q_s \over 24 \pi f^2} \,{Q^3 \over \sqrt{t}}\,
(3F+D)^2\, {\mns \over P^2}\\
&&\left\{ \left({{\bar M}^2 \over P Q}\right) 
Q_2(\xbb) +\left({{\bar M}^2 \Delta{\tilde M} \over Q^2}
\right) Q_1(\xbb)\right\} \nonumber \ .
\end{eqnarray}
Up to the kaon strangeness form factor, these expressions 
are exactly the same as
obtained by the Born approximation for the $\bll$
and the dispersion relation approach (compare with Eqs. 
(\ref{foba},\ref{ftba})).
The imaginary parts Eqs. (\ref{specfunc1}), (\ref{specfunc2}) 
are defined for $t \geq 4 m_N^2$.
Since the discontinuity starts at $4 m_K^2$, they
have to be analytically continued 
into the unphysical region $4 m_K^2 \leq t < 4 m_N^2$. 
This is easily done by replacing the momentum
$P = \sqrt{t/4 -m_N^2}$ by $i\,p_{-} = i\sqrt{m_N^2 -t/4}$.
Consequently, the variable $\xbb$ becomes complex ($\xbb \to -i \xi_B$),
and the Legendre, functions
of the second kind have to be analytically continued, too.

\begin{figure}
\epsfysize=5in
\rotate[r]{\epsffile{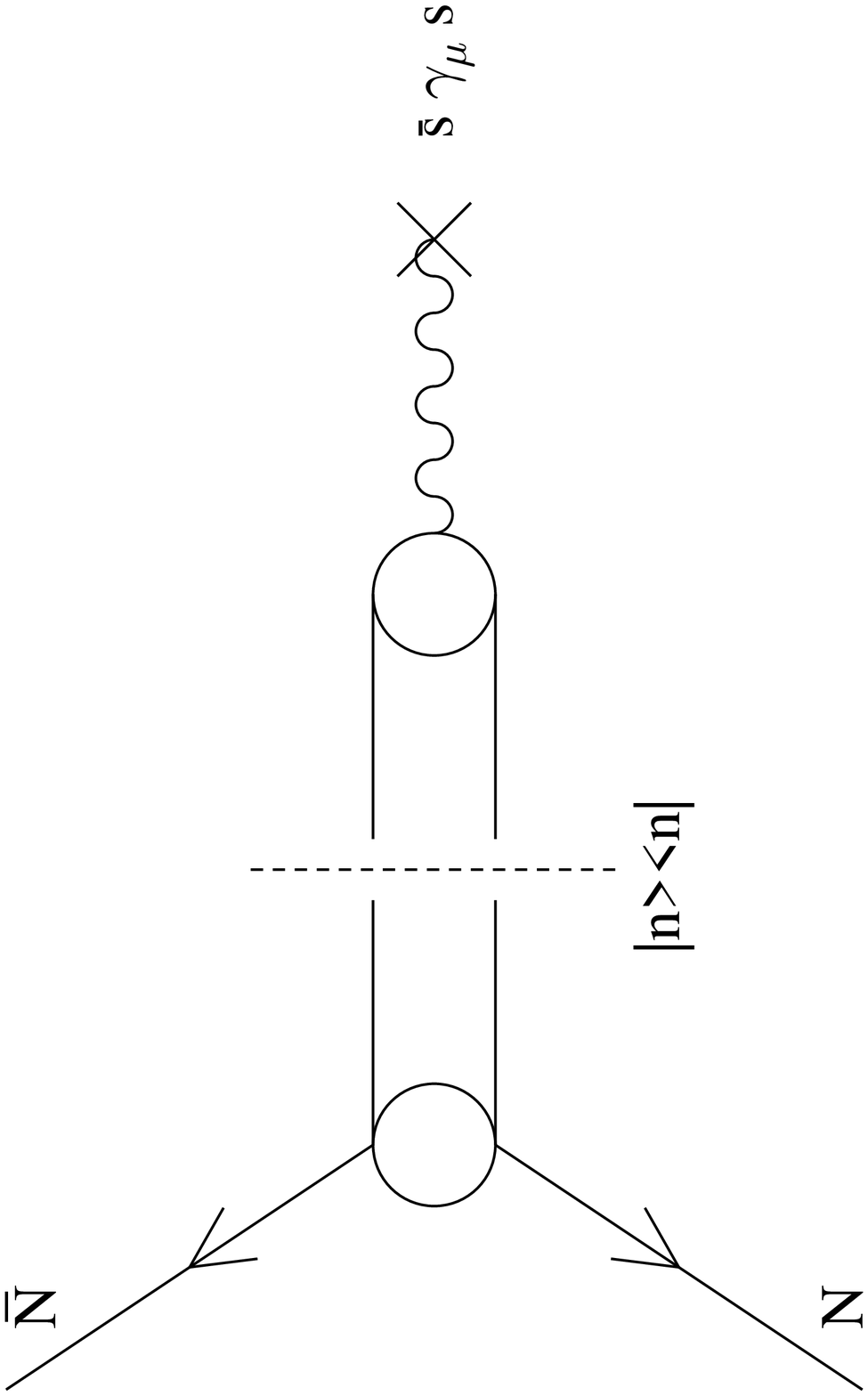}}
\caption{\label{Fig1} Diagrammatic representation of the
spectral decomposition for the nucleon's strangeness
vector current form factors given in Eq. (\ref{spec_t}). Right
hand part of the diagram denotes the matrix element to
produce a $I^G(J^{PC})=0^-(1^{--})$ state from the vacuum
through the strangeness vector current. Left hand side
denotes the $n\to N\bar{N}$ scattering amplitude. }
\end{figure}

\begin{figure}
\epsfysize=5in
\rotate[r]{\epsffile{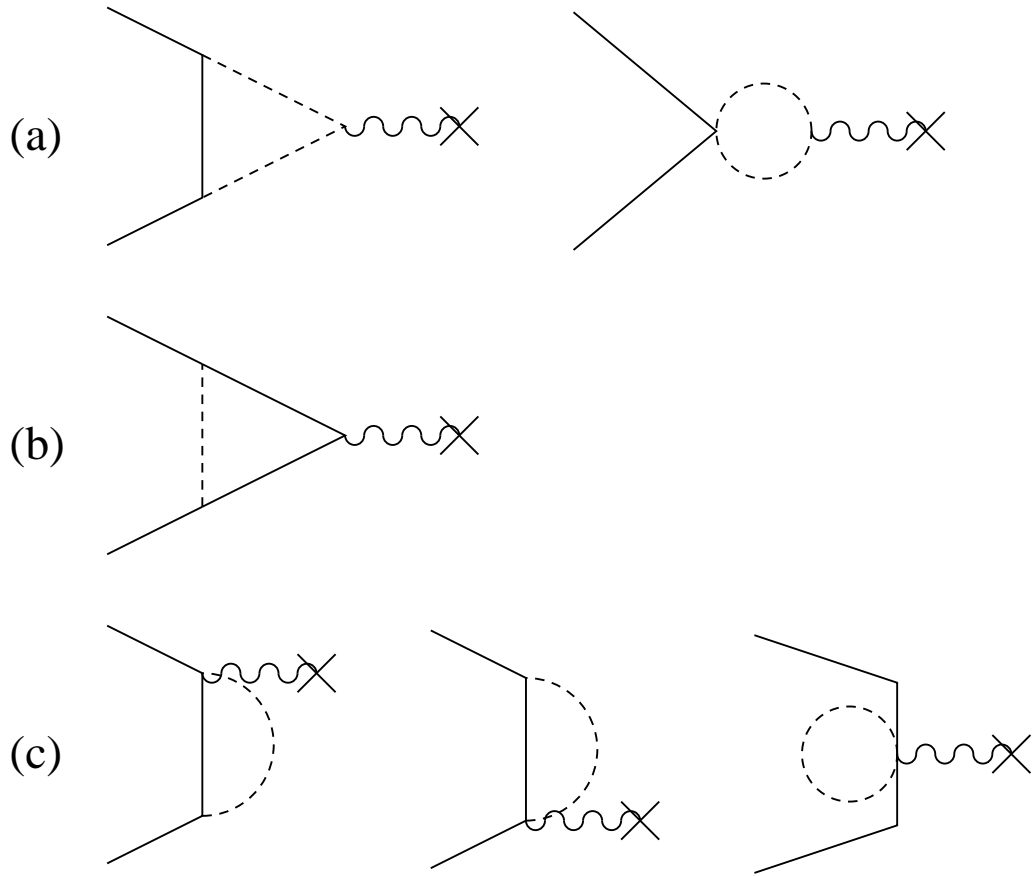}}
\caption{\label{Fig2} One-loop diagrams for the strange
vector form factors of the nucleon; the strange vector current
$\bar{s}\gamma_\mu s$ is denoted by the curly line, the 
dashed lines correspond to kaons, and the solid lines
correspond to nucleons (external) or strange baryons
(internal to loop).}
\end{figure}

\begin{figure}
\epsfysize=5in
\rotate[r]{\epsffile{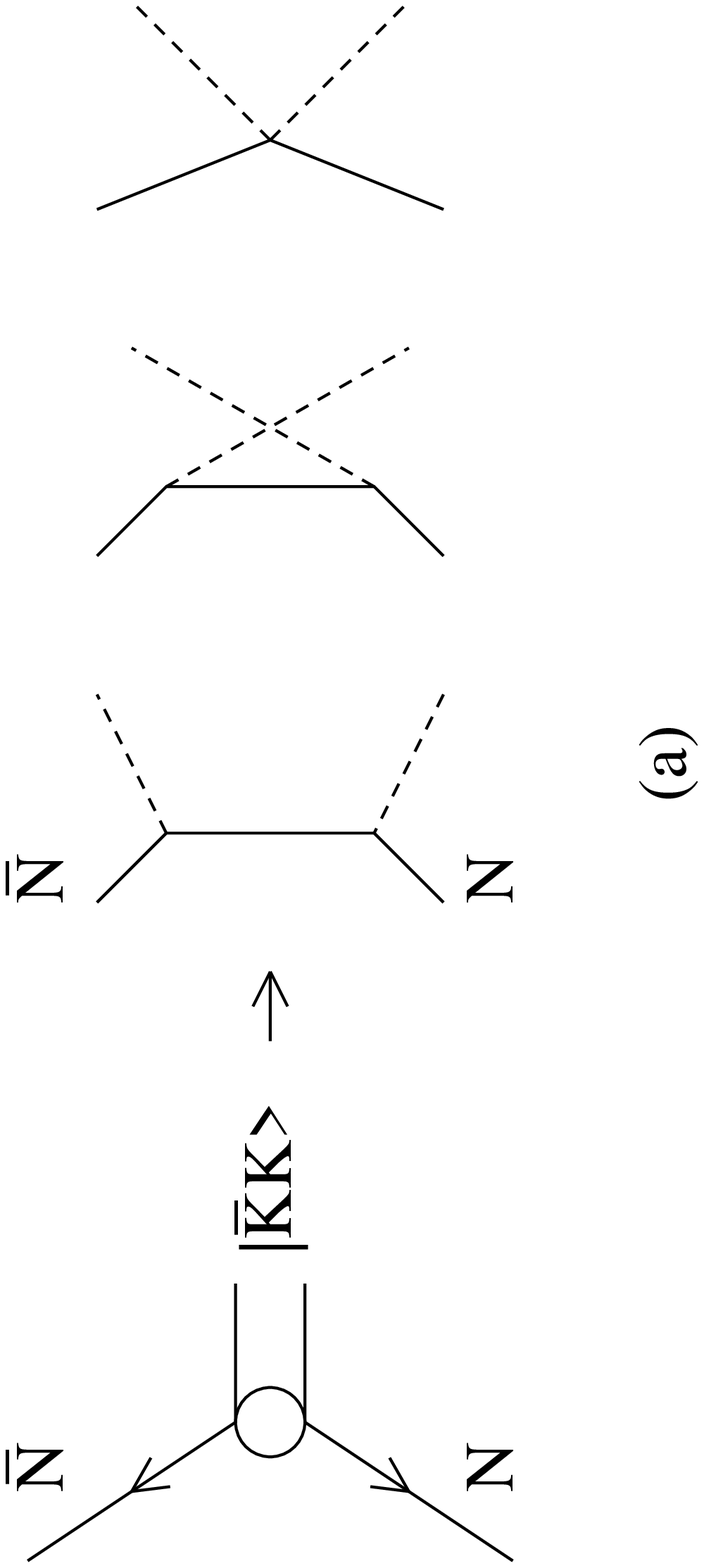}}
\caption{\label{Fig3} Approximations for the $n\to N\bar{N}$
scattering amplitude appearing in Fig. \protect\ref{Fig1} 
and Eq. (\ref{spec_t}).
Panel (a) gives Born approximation for the $K\bar{K}\to N\bar{N}$
amplitude, while (b) represents the $B\bar{B}\to N\bar{N}$ amplitude
in the one meson-exchange approximation ($B$ is a baryon).}
\epsfysize=5in
\rotate[r]{\epsffile{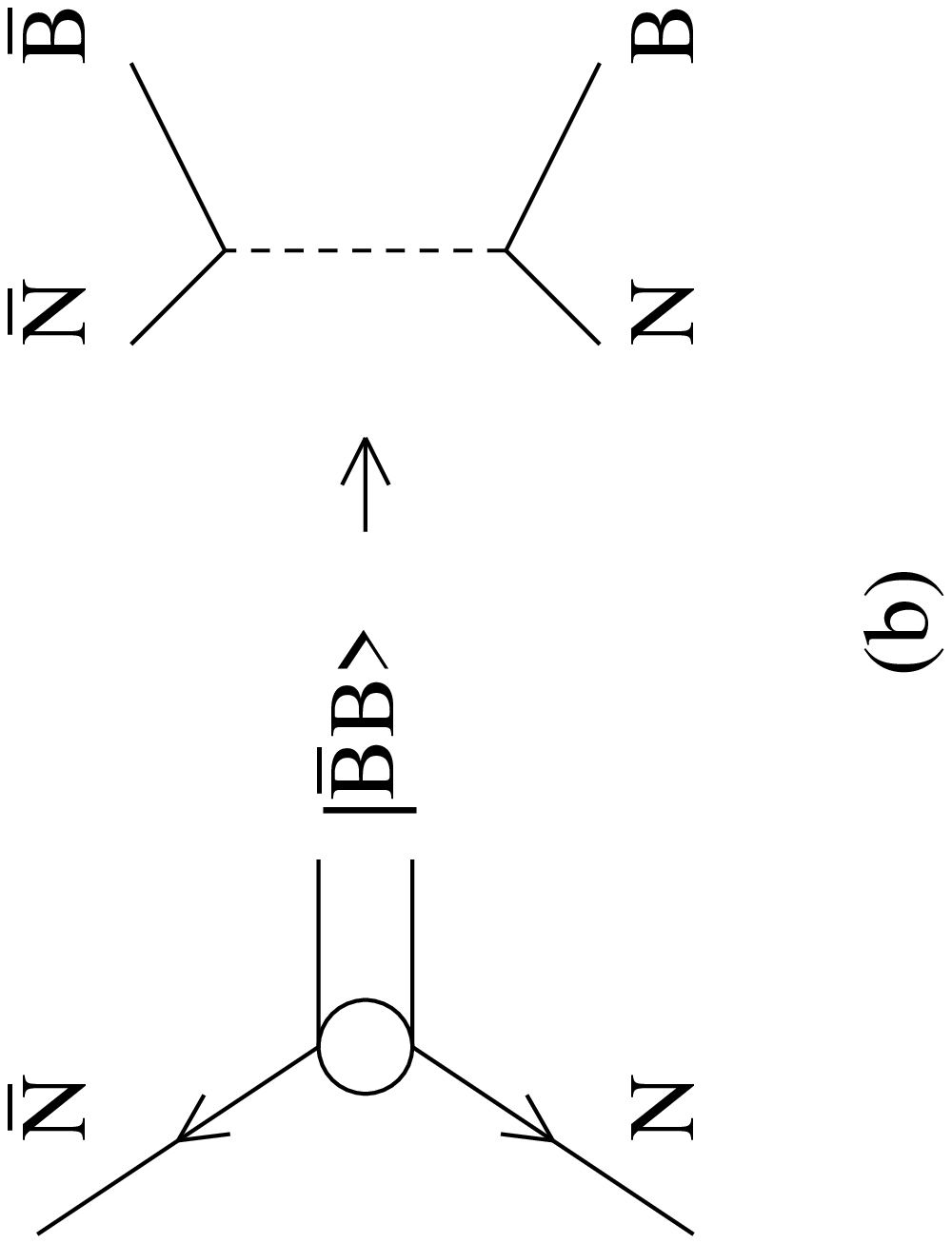}}
\end{figure}

\begin{figure}
\epsfysize=5in
\rotate[r]{\epsffile{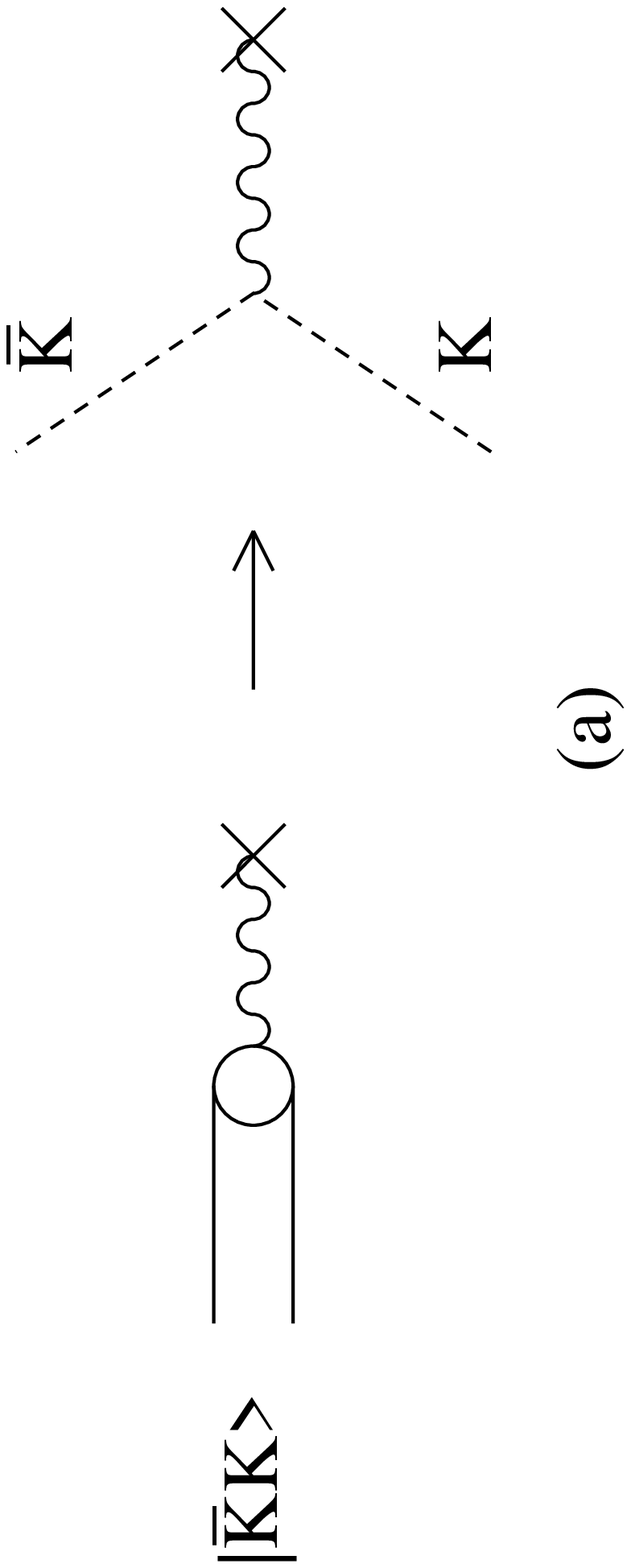}}
\caption{\label{Fig4} Pointlike approximation for the matrix elements
$\bra{n}\bar{s}\gamma_\mu s\ket{0}$ entering the spectral functions
as in Eq. (\ref{spec_t}) and Fig. \protect\ref{Fig1}. Panel (a) corresponds
to pointlike kaon strangeness form factor, while (b) denotes the same
for a baryon.}
\epsfysize=5in
\rotate[r]{\epsffile{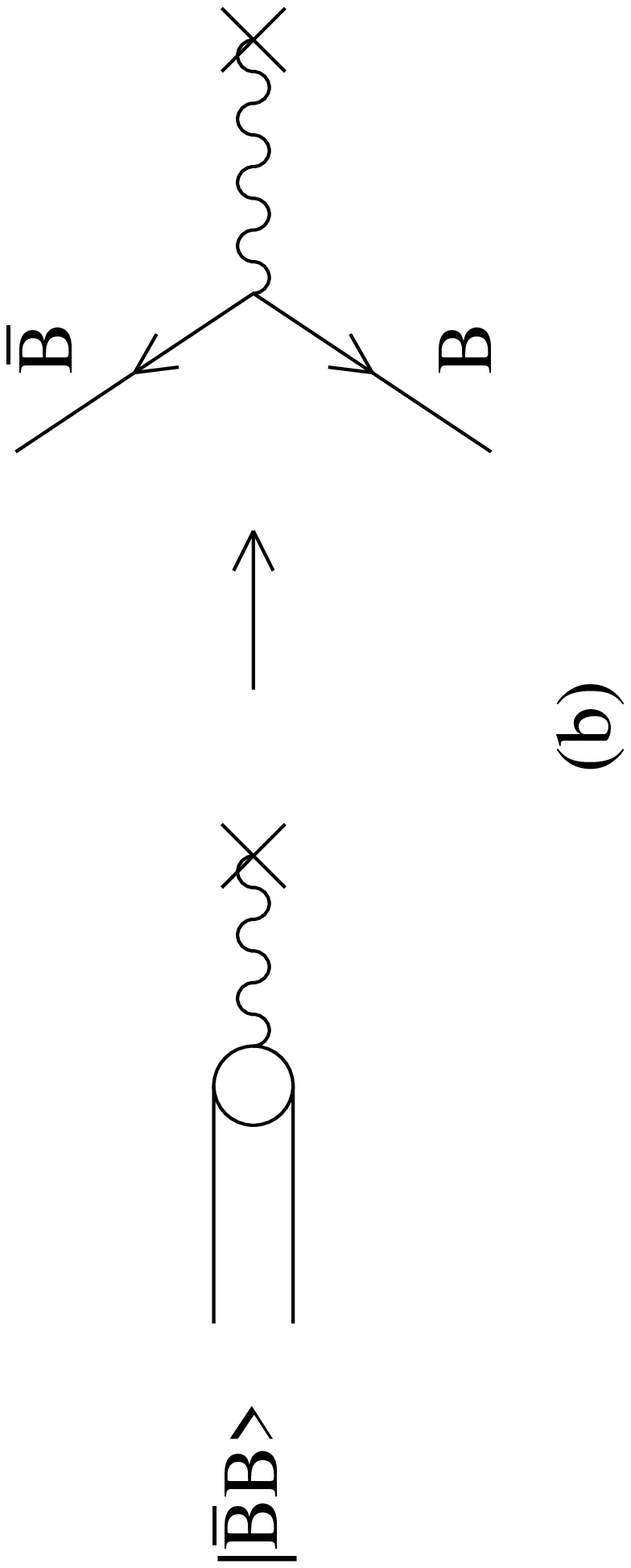}}
\end{figure}

\begin{figure}
\epsfysize=5in
\rotate[r]{\epsffile{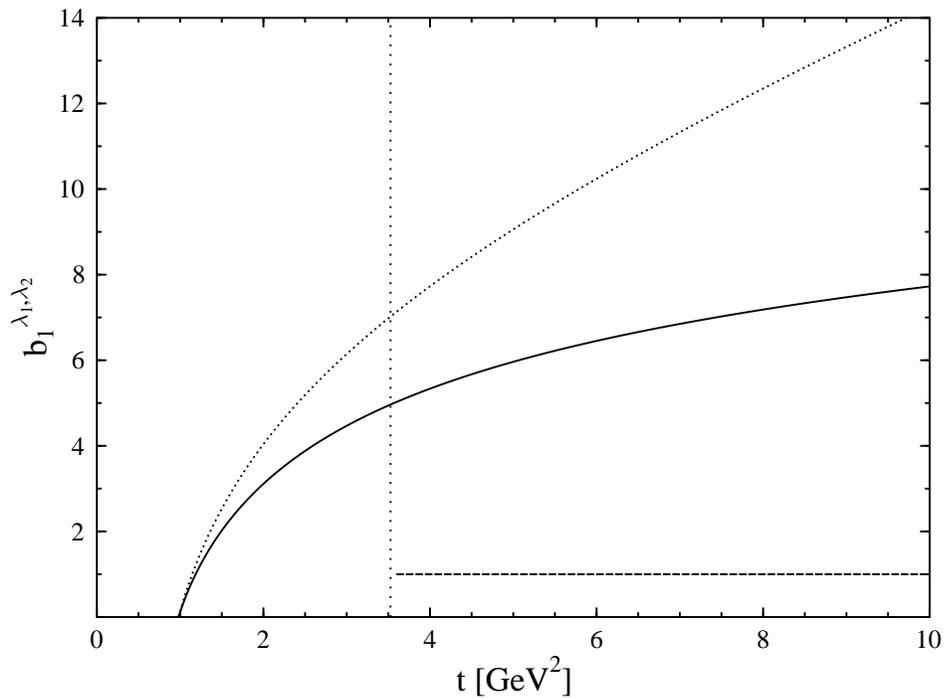}}
\caption{\label{Fig5} Partial waves $\bll$ for $KN$-scattering in the 
non-linear SU(3) $\sigma$-model. The solid and dotted
lines correspond to $\bpp$ and $\bpm$, respectively. The dashed line
shows the unitarity bound on $\bpm$; the bound on
$\bpp$, which is not shown, is a factor $1/\sqrt{2}$ smaller
at $N\bar{N}$ threshold, indicated by the vertical dotted line.}
\end{figure}

\begin{figure}
\epsfysize=5in
\rotate[r]{\epsffile{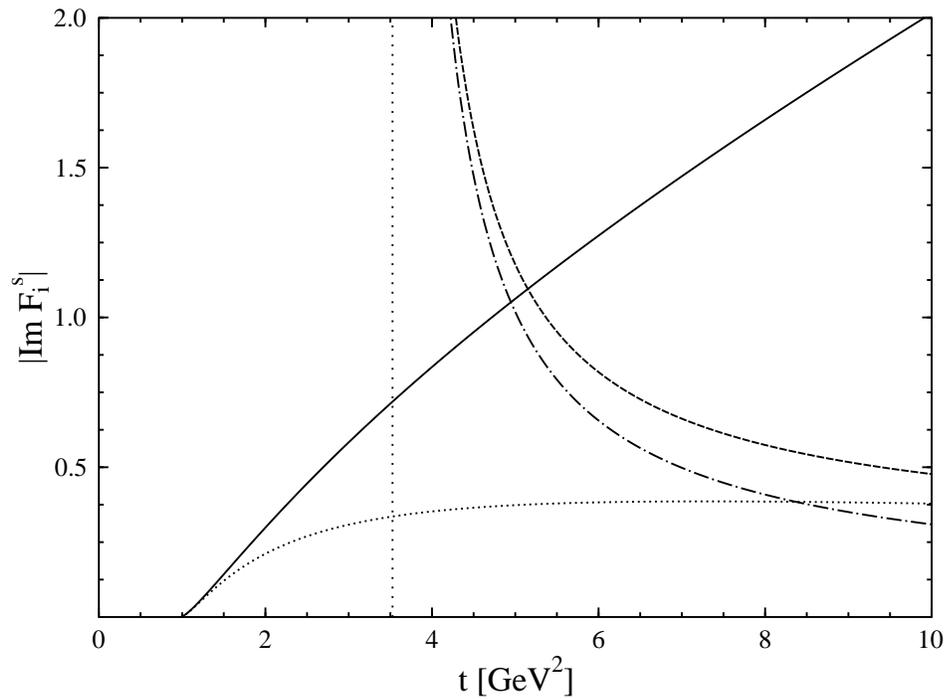}}
\caption{\label{Fig6} Spectral functions in the non-linear
$\sigma$-model and na\"\i ve unitarity bounds.
A point-like strangeness form factor for the kaon has been used. 
The solid and dotted lines show the results for $\hbox{Im}\ \FOS(t)$
and $\hbox{Im}\ \FTS(t)$, respectively. The corresponding
na\"\i ve unitarity bounds are indicated by the 
dashed and dash-dotted lines, respectively.
The vertical dotted line indicates the two-nucleon threshold.}
\end{figure}

\begin{figure}
\epsfysize=5in
\rotate[r]{\epsffile{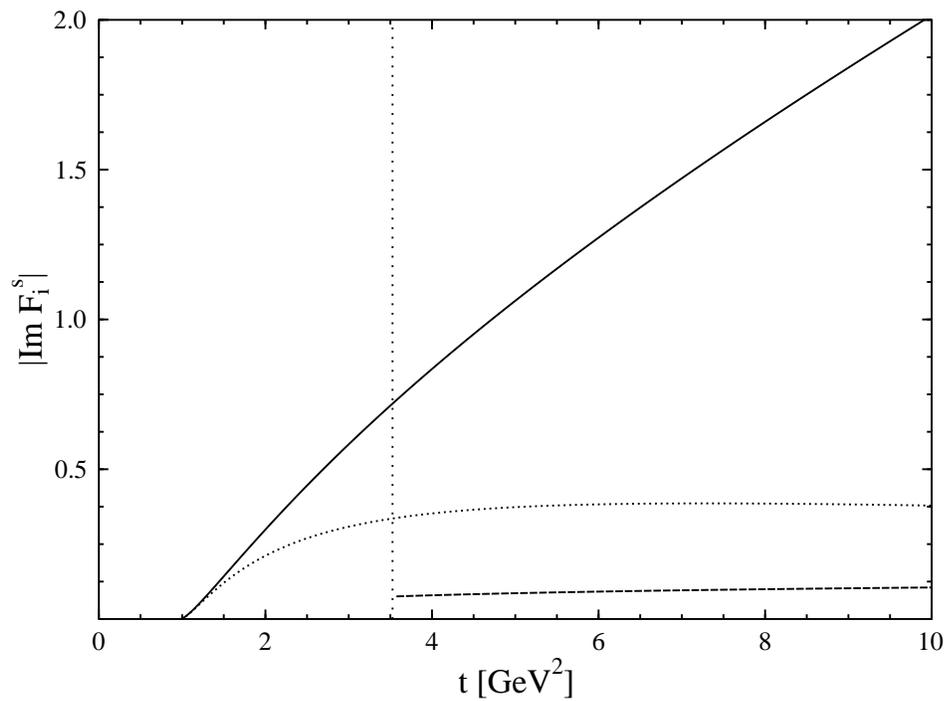}}
\caption{\label{Fig7} Same as Fig. \protect\ref{Fig6}
but with the {\it correct} unitarity bounds of Eqs. (\ref{ubf1th}, 
\ref{ubf2th}).
The bound on $\hbox{Im}\ \FTS(t)$ is not displayed, because
it is even more stringent than the bound on $\hbox{Im}\ \FOS(t)$.}
\end{figure}

\begin{figure}
\epsfysize=5in
\rotate[r]{\epsffile{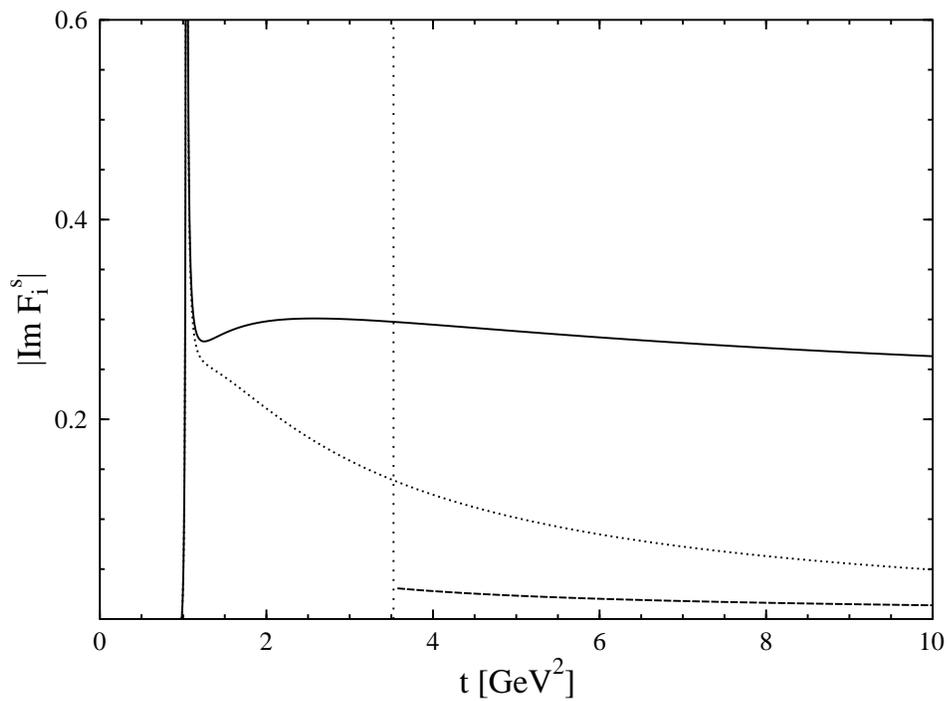}}
\caption{\label{Fig8} Same as Fig. \protect\ref{Fig7}
but using the GS parametrization for the kaon strangeness form
factor, peaked for $\sqrt{t}\approx\mphi$. Note the difference
in vertical scale as compared to Figs. \protect\ref{Fig6} and 
\protect\ref{Fig7}.}
\end{figure}

\begin{figure}
\epsfysize=5in
\rotate[r]{\epsffile{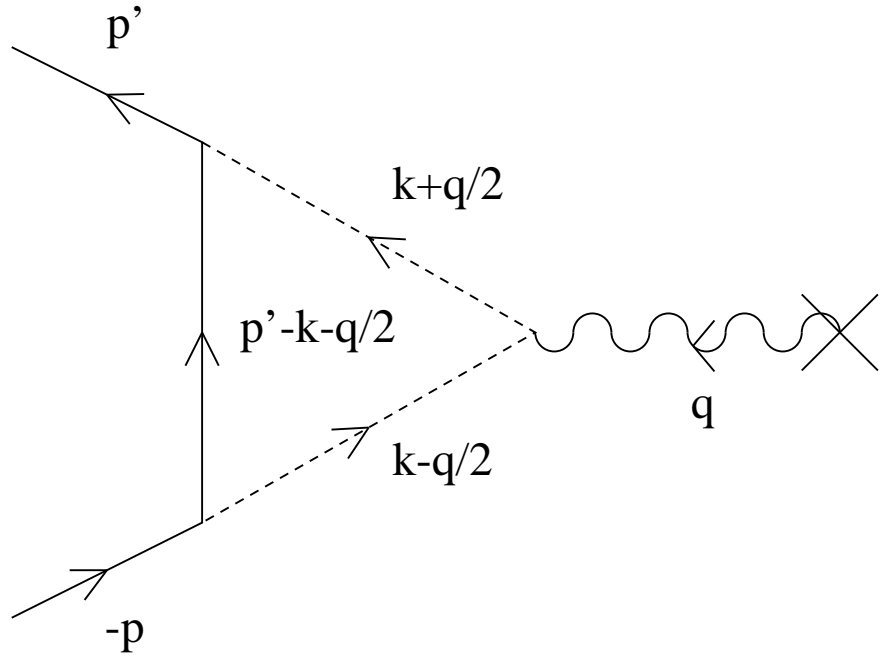}}
\caption{\label{Fig9} Our choice of the internal and external momenta
for the calculation of the imaginary parts 
arising from the t-channel discontinuity of the diagrams
from Fig. \protect\ref{Fig2}a .}
\end{figure}


\begin{references}
\bibitem{Ste96} Since the reaction $K\bar{K}\to N\bar{N}$ is an inelastic
		one, a partial wave expansion of the $S$-matrix element 
		is equivalent to a partial wave expansion of the 
		corresponding $T$-martrix element, up to constant factors.
\bibitem{com1}  In the present context, we avoid making the heavy baryon
		expansion of the non-linear chiral lagrangian, even
		though such an expansion provides a well-defined
		truncation scheme for baryon CHPT with loops. For
		$KN$ scattering in the B.A., however, the heavy baryon
		lagrangian does not yield the correct pole structure.
\end{references}
\end{document}